\documentclass[12pt]{article}
\usepackage{amsfonts,amsmath,graphicx,setspace}
\usepackage{color,hyperref,verbatim,natbib,rotating}

\usepackage{booktabs} 
\usepackage[labelfont=bf]{caption} 

\usepackage{lscape}

\usepackage{placeins}
\usepackage{float}

\usepackage{adjustbox}

\usepackage{xcolor} 
\definecolor{lightgray}{rgb}{0.8, 0.8, 0.8} 
\usepackage{colortbl} 
\newcolumntype{g}{>{\columncolor{lightgray!40}}c} 

\newcommand{\tildeL}[1]{\stackrel{\sim}{\smash{#1}\rule{0pt}{1.5ex}}} 
\newcommand{\indep}{\perp \!\!\! \perp} 

\definecolor{Red}{rgb}{0.5,0,0}
\definecolor{Blue}{rgb}{0,0,0.5}
\hypersetup{%
    colorlinks = {true},
    linktocpage = {true},
    plainpages = {false},
    linkcolor = {Blue},
    citecolor = {Blue},
    urlcolor = {Red},
    pdfstartview = {XYZ null null 1.25},
    pdfpagemode = {UseOutlines},
    pdfview = {XYZ null null null}
}
\newcommand{\email}[1]{\href{mailto:#1}{\normalfont\texttt{#1}}}

\bibpunct{(}{)}{;}{a}{}{,}

\begin{document}


\begin{center}
\singlespacing\Large \bf A joint model for (un)bounded longitudinal markers, competing risks, and recurrent events using patient registry data
\end{center}


\begin{center}
{\large Pedro Miranda Afonso$^{1,2,\ast}$, Dimitris Rizopoulos$^{1,2}$, Anushka K. Palipana$^{3,4}$, Emrah Gecili$^{4,5}$, Cole Brokamp$^{4,5}$, John P. Clancy$^{6}$, Rhonda D. Szczesniak$^{4,5,7}$ and Eleni-Rosalina Andrinopoulou$^{1,2}$}\footnote{$^\ast$Correspondence at: Department of Biostatistics, Erasmus University Medical Center, PO Box 2040, 3000 CA Rotterdam, the Netherlands. E-mail address: \email{p.mirandaafonso@erasmusmc.nl}.}\\
\end{center}
\singlespacing{
$^{1}$Department of Biostatistics, Erasmus University Medical Center, the Netherlands; $^{2}$Department of Epidemiology, Erasmus University Medical Center, the Netherlands, $^{3}$School of Nursing, Duke University, USA; $^{4}$Division of Biostatistics and Epidemiology, Cincinnati Children’s Hospital Medical Center, USA; $^{5}$Department of Pediatrics, University of Cincinnati, USA; $^{6}$Division of Statistics and Data Science, University of Cincinnati, USA; $^{7}$Division of Pulmonary Medicine, Cincinnati Children’s Hospital Medical Center, USA\\
}


\begin{spacing}{1}
\noindent {\bf Abstract}\\
Joint models for longitudinal and survival data have become a popular framework for studying the association between repeatedly measured biomarkers and clinical events. Nevertheless, addressing complex survival data structures, especially handling both recurrent and competing event times within a single model, remains a challenge. This causes important information to be disregarded. Moreover, existing frameworks rely on a Gaussian distribution for continuous markers, which may be unsuitable for bounded biomarkers, resulting in biased estimates of associations. To address these limitations, we propose a Bayesian shared-parameter joint model that simultaneously accommodates multiple (possibly bounded) longitudinal markers, a recurrent event process, and competing risks. We use the beta distribution to model responses bounded within any interval $(a,b)$ without sacrificing the interpretability of the association. The model offers various forms of association, discontinuous risk intervals, and both gap and calendar timescales. A simulation study shows that it outperforms simpler joint models. We utilize the US Cystic Fibrosis Foundation Patient Registry to study the associations between changes in lung function and body mass index, and the risk of recurrent pulmonary exacerbations, while accounting for the competing risks of death and lung transplantation. Our efficient implementation allows fast fitting of the model despite its complexity and the large sample size from this patient registry. Our comprehensive approach provides new insights into cystic fibrosis disease progression by quantifying the relationship between the most important clinical markers and events more precisely than has been possible before. The model implementation is available in the \textsf{R} package \texttt{JMbayes2}.
\end{spacing}



\section{Introduction} \label{sec:intro}

In clinical research, joint models for longitudinal and survival data have become a popular framework for studying biomarkers measured over time and their association with clinical events~\citep{henderson2000joint, tsiatis2004joint, rizopoulos2012joint}. Several extensions have been developed to the basic framework for a single event time and a continuous longitudinal biomarker proposed by \cite{faucett1996simultaneously} and \cite{wulfsohn1997joint}. The literature is extensive, with recent comprehensive reviews by \cite{hickey2016joint, hickey2018joint}, \cite{papageorgiou2019overview}, and \cite{alsefri2020bayesian}. These reviews reflect the ongoing efforts to enhance the versatility of the framework and its ability to address the intricate features often found in longitudinal and survival data. 

Cystic fibrosis (CF) is a severe genetic disorder that primarily affects the lungs and digestive system, leading to respiratory impairment and malnutrition~\citep{farrell2008guidelines}. Patients with CF often experience recurrent lung infections, known as pulmonary exacerbations (PEx), which can cause permanent lung damage and increase the risks of lung transplantation and death. The body mass index (BMI) and the percentage of predicted forced expiratory volume in one second (ppFEV\textsubscript{1}) are routinely measured to monitor disease progression. CF care teams are interested in understanding the associations between ppFEV\textsubscript{1} decline, BMI changes, recurrent PEx, and the competing risks of death and lung transplantation using the US Cystic Fibrosis Foundation Patient Registry (CFFPR,~\citealp{knapp2016cystic}). Previous studies that aimed to investigate such associations using joint models were hampered by the lack of an appropriate framework. 

The joint modeling framework has previously been extended to incorporate complex survival data structures, such as recurrent \citep{liu2008analysis, liu2009joint, kim2012joint, krol2016joint} and competing event time data \citep{ elashoff2008joint, williamson2008joint,andrinopoulou2014joint}. However, the integration of both recurrent events and competing risks within a unified model remains a challenge, leading researchers to omit important information despite availability in patient registries. For example, \cite{andrinopoulou2020multivariate} limited their analysis to the period up to the first PEx event, disregarding subsequent occurrences and informative censoring due to transplantation or death. When investigating the association between ppFEV\textsubscript{1} and the risks of death and lung transplantation, \cite{afonso2023efficiently} treated these two events as a composite endpoint rather than as competing risks, assuming that they indicate the same prior health status, which is not clinically accurate. 

An additional limitation of existing frameworks is their tendency to rely exclusively on the Gaussian distribution to model continuous markers. An important aspect of joint modeling is the appropriate parameterization of longitudinal submodels to ensure accurate extrapolation of unobserved biomarker evolution up to the event time. A Gaussian parameterization can be problematic for a bounded biomarker with many observations close to the boundaries, such as ppFEV\textsubscript{1}, as it can cause the model to yield biologically implausible values, resulting in biased estimates of the marker evolution and its associations. Existing CF studies have modeled ppFEV\textsubscript{1} mostly using a Gaussian distribution. \cite{szczesniak2023lung} explored the use of other distributions; however, deriving a meaningful clinical interpretation from the association in the linear predictor scale was challenging.

We address these collective limitations by introducing a comprehensive joint modeling framework that can (i)~effectively accommodate competing risk and recurrent event processes together with multiple longitudinal outcomes, and (ii)~appropriately model bounded longitudinal markers with constrained distributions, without compromising the interpretability of their association. Our model captures the complex dynamics of CF by simultaneously considering recurrent PEx and the competing risks of death and lung transplantation, and by appropriately parameterizing the longitudinal markers ppFEV\textsubscript{1} and BMI using beta and Gaussian distributions, respectively. The choice of a beta distribution ensures that ppFEV\textsubscript{1} remains within the feasible range. The model allows for the use of various functional forms to link time-to-event and longitudinal processes, and accommodates discontinuous risk intervals and both gap and calendar timescales. The model has been made available in the user-friendly \textsf{R} package for joint models, \texttt{JMbayes2}~\citep{jmbayes2}, which is available in the Comprehensive \textsf{R} Archive Network (CRAN). The implementation approach emphasized versatility and efficiency to streamline the package's adoption in complex settings with large sample sizes.

The remainder of this article is organized into four sections. Section~\ref{sec:meth} describes the proposed joint modeling framework in detail. In Section~\ref{sec:sim}, a simulation study demonstrates the added value of our approach over simpler joint models. In Section~\ref{sec:app}, we apply the proposed model in a real-world setting using the CFFPR dataset. Lastly, Section~\ref{sec:discuss} summarizes the main findings and outlines directions for future research.

\section{Joint modeling framework} \label{sec:meth}

We propose a joint model with $J$ longitudinal markers that can follow different distributions, $K$ competing events, and one recurrent event process.  Joint models assume a full joint distribution of the longitudinal and time-to-event processes that can be factorized in different ways~\citep{sousa2011review}. We focus on the shared-parameter joint models in this work; we assume that the time-to-event and longitudinal processes depend on an unobserved process defined by random effects. The observed processes are assumed independent conditional on the random effects. Below we present the submodels that make up the proposed joint model.

\subsection{Longitudinal outcomes} \label{sec:meth:long}

To describe the subject-specific time evolution of the $j$th longitudinal outcome, we consider a mixed-effects regression model
\begin{equation*}\label{eq:meth:long}
\begin{cases}
\textbf Y_{j,i} \mid \textbf b_{j,i} \sim \mathcal{F}_{j,\Psi_j}\\\
\textbf b_{j,i} \sim \mathcal{N}\left(\textbf 0, \textbf D_j\right),
\end{cases}
\end{equation*}
where $\textbf Y_{j,i}$ is the $j$th response for the $i$th individual, $\textbf b_{j,i}$ is the corresponding vector of random effects and $\mathcal F_j$ is a set of discrete and continuous distributions (not restricted to the exponential family). The random effects follow a zero-mean multivariate normal distribution with unstructured variance-covariance matrix $\textbf D_j$. The expected value of the $j$th outcome at time $t$ conditional on the random effects, $\mu_{j,i}(t)=\text E\{\text Y_{j,i}(t) \mid \textbf b_{j,i}\}$, has the form
\begin{equation}\label{eq:mth:meff}
\mu_{j,i}(t) = \mathcal{G}^{-1}_j\left\{\eta_{j,i}(t)\right\}= \mathcal{G}^{-1}_j\left\{\textbf x^\top_{j,i}(t)\boldsymbol\beta_j + \textbf z^\top_{j,i}(t)\textbf b_{j,i}\right\},
\end{equation}
where $\eta_{j,i}(t)$ is the linear predictor, $\textbf x_{j,i}(t)$ and $\textbf z_{j,i}(t)$ are the design vectors of (possibly time-varying) covariates for the fixed effects $\boldsymbol{\beta}_j$ and the subject-specific random effects $\textbf{b}_{j,i}$, respectively, and $\mathcal{G}_j(\cdot)$ is the link function. In this work, given the motivating case study, we focus our attention on two particular continuous distributions: Gaussian and beta.  

Let $Y_{j,i}(t)$ be a random sample drawn from the distribution $\text{Beta}\left(p,q\right)$ with nonnegative shape parameters $p$ and $q$. We follow the beta density reparameterization proposed by~\cite{ferrari2004beta}, which is indexed by the mean $\mu_{j,i}=p/(p+q)$ and a precision parameter $\phi=p+q$, which satisfies $0<\mu_{j,i}(t)<1$ and $\phi>0$. This choice stems from the difficulty of interpreting shape parameters in terms of conditional expectations. The flexibility of the beta density enables it to adopt a plethora of distinctive shapes ranging from symmetric bell-shaped curves to flat, skewed, or U-shaped curves within the open interval $(0, 1)$~\citep{gupta2004handbook}. This versatility makes the beta distribution an appealing choice for modeling a continuous outcome that takes values within a known interval, such as in the case of ppFEV\textsubscript{1}. We focus on the logit link $\log\{\mu/(1-\mu)\}$ in this work, but other link functions can be used. For the logit link, the submodel's regression parameters $\boldsymbol{\beta}_j$ are interpretable in terms of expected changes in $\mbox{logit}\{\mu_{j, i}(t)\}$. Effects plots can be employed to retrieve these interpretations to the original scale.

The model is heteroscedastic because the variance of $Y_{j,i}(t)$ is a function of its expected value, $\mbox{Var}\left\{Y_{j,i}(t)\right\}=\mu_{j,i}(t)\{1-\mu_{j,i}(t)\}/(1+\phi)$. Thus, the model intrinsically accommodates non-constant response variances.

When considering a normally distributed outcome, we use the identity link function in~\eqref{eq:mth:meff}, such that $\mu_{j,i}(t)=\eta_{j,i}(t)$, and we account for the measurement error by including the term $\varepsilon_{j,i}(t)$ in $Y_{j,i}(t)=\eta_{j,i}(t)+\varepsilon_{j,i}(t)$, where $\varepsilon_{j,i}(t)\sim\mathcal{N}(0,\sigma^2_{\text{y}_j})$. We assume the measurement errors $\varepsilon_{j,i}(t)$ to be mutually independent and independent of the random effects $\textbf b_{j,i}$. Multiple longitudinal outcomes are associated through the variance-covariance matrix $\textbf{D}$, which encompasses the $J$ variance-covariance matrices $\textbf{D}_j$. Joint models using the Gaussian distribution have been extensively discussed in the literature (see, for example, \citealp{rizopoulos2014combining}).

\subsection{Recurrent event times} \label{sec:meth:rec}

For the risk of the recurring event, we rely on a proportional hazards risk model. The hazard function for the $l$th event at time $t$ is modeled by 
\begin{equation*}\label{eq:meth:rec}
h^{\mbox{\tiny\text{R}}}_i(t)= h^{\mbox{\tiny\text{R}}}_0\left(t-t_{0_{l,i}}\right)\exp \Big[  \textbf w^{\mbox{\tiny\text{R}}\top}_i(t) \boldsymbol\gamma^{\mbox{\tiny\text{R}}}
+\sum_{j=1}^{J}\sum_{m=1}^{M_j}\mathcal{H}^{\mbox{\tiny\text{R}}}_{j,m}\left\{\eta_{j,i}(t)\right\}\alpha^{\mbox{\tiny\text{R}}}_{j,m}  + \upsilon^{\mbox{\tiny\text{R}}}_i\Big], 
\end{equation*}
for $t>t_{0_{l,i}}\geq0$, where $t_{0_{l,i}}$ is the starting time of the risk interval for the $l$th recurrent event, and $\upsilon^{\mbox{\tiny\text{R}}}_i \sim \mathcal{N}\left(0, \sigma_\upsilon^2\right)$. For the baseline hazard function $h^{\mbox{\tiny\text{R}}}_0\left(t-t_{0_{l,i}}\right)$, we use penalized B-spline functions, P-splines~\citep{eilers1996flexible}. Specifically, we use $\log h_0^{\mbox{\tiny\text{R}}}\left(t-t_{0_{l,i}}\right)=\sum_{q=1}^{Q}\gamma^{\mbox{\tiny\text{R}}}_{0_q}\text{bs}^{\mbox{\tiny\text{R}}}_q\left(t-t_{0_{l,i}}\right)$, where $\text{bs}^{\mbox{\tiny\text{R}}}_q(t)$ are the P-splines' $q$th basis functions of degree $d$, and $\gamma^{\mbox{\tiny\text{R}}}_{0_q}$ are the corresponding unknown coefficients. In the relative risk component of the model, the design vector $\textbf w^{\mbox{\tiny\text{R}}}_i(t)$ contains the measured characteristics with the corresponding vector of regression coefficients $\boldsymbol{\gamma}^{\mbox{\tiny\text{R}}}$; the design vector may incorporate baseline or time-varying exogenous covariates.

The hazard of an event for individual $i$ at time $t$ is associated with the $j$th subject-specific marker trajectory through the latent association structure $\mathcal{H}^{\mbox{\tiny\text{R}}}_{j,m}\left\{\eta_{j,i}(t)\right\}=\mathcal{H}^{\mbox{\tiny\text{R}}}_{j,m}\left\{\eta_{j,i}(u)\right\}$, $0\leq u\leq t$, which include the random effects $\textbf b_{j,i}$. The longitudinal and recurrent event processes are assumed to be conditionally independent given $(\textbf b^\top_{1,i},\cdots,\textbf b^\top_{J,i})$. The function $\mathcal{H}^{\mbox{\tiny\text{R}}}_{j,m}(\cdot)$ determines the form of association between the longitudinal and time-to-event processes. The available functional forms are elaborated upon in Section~\ref{sec:meth:forms}. The association parameter $\alpha^{\mbox{\tiny\text{R}}}_{j,m}$ measures the strength of the association between the $m$th functional form of the $j$th longitudinal outcome and the risk of the next event. The quantity $\exp\left\{\alpha^{\mbox{\tiny\text{R}}}_{j,m}\right\}$ is the hazard ratio (HR) for a one-unit increase in the value of $\mathcal{H}^{\mbox{\tiny\text{R}}}_{j,m}\left\{\eta_{j,i}(t)\right\}$ while the rest of the variables are kept constant.

We incorporate the random effect $\upsilon^{\mbox{\tiny\text{R}}}_i$ to capture the correlation among event times within the same individual. Hereafter, we refer to the random effect terms in the risk models as frailties to distinguish them from the random effects in the longitudinal submodels. We assume that the subject-specific frailties and random effects are independent of each other, and that the event times from the same individual are independent conditional on $\upsilon_i^{\mbox{\tiny\text{R}}}$.

\begin{figure}
\centerline{\includegraphics[width=0.9\textwidth]{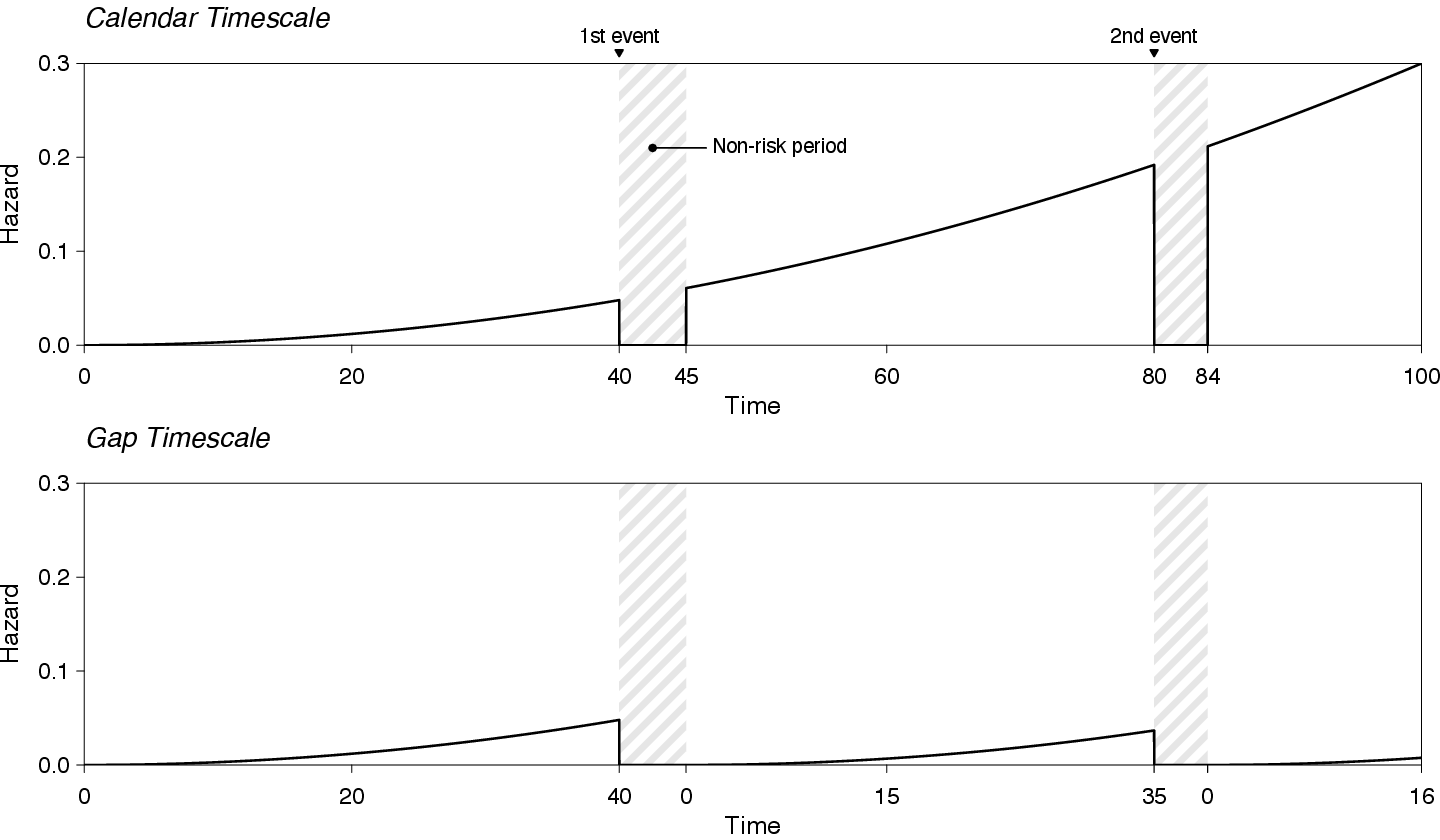}}
\caption{The hazard function for a hypothetical recurrent event process, assuming the calendar (top~panel) or gap (bottom~panel) timescale. During the study period, from time 0 to 100, the displayed individual experienced two recurrent events (e.g., hospitalizations) at times 40 and 80. These events lasted five and four time units, respectively; during these periods, the individual was not at risk of a new event.}
\label{fig:mth:timescale}
\end{figure}

Our approach allows the recurrent event process to be modeled under the gap or calendar timescales, which use different zero-time references~\citep{duchateau2003evolution}. As shown in the illustrative example in Figure~\ref{fig:mth:timescale}, the calendar timescale uses a shared reference time for all events (e.g., study entry), $t_{0_{l,i}}=0, \forall\;l$, while the gap timescale uses the end of the previous event, assuming a renewal after each event and resetting the time to zero. Furthermore, our model accommodates non-risk periods in which a patient is still experiencing the previous event and so is not yet at risk of experiencing the next one. For example, if we are interested in modeling the time to the next hospitalization, then a patient who is currently hospitalized is not at risk of being hospitalized again.

\subsection{Competing risks} \label{sec:meth:term}

To model the risks associated with each of the competing events, we consider a cause-specific hazard, allowing for distinct specific forms of association between the longitudinal outcomes and each cause of failure. The instantaneous rate for failures of cause $k$ at any time $t>0$ is modeled by
\begin{equation*}\label{eq:meth:term}
h^{\mbox{\tiny\text{T}}}_{k,i}(t)= h^{\mbox{\tiny\text{T}}}_{0_k}(t)\exp \Big[\textbf w^{\mbox{\tiny\text{T}}\top}_{k,i}(t) \boldsymbol \gamma^{\mbox{\tiny\text{T}}}_k+
+ \sum_{j=1}^{J}\sum_{m=1}^{M_j} \mathcal{H}^{\mbox{\tiny\text{T}}}_{k,j,m}\left\{\eta_{j,i}(t)\right\}\alpha^{\mbox{\tiny\text{T}}}_{k,j,m} + \upsilon^{\mbox{\tiny\text{T}}}_{k,i}\Big],
\end{equation*}
by censoring all other causes. Here, $h^{\mbox{\tiny\text{T}}}_{0_k}(t)$ is the cause-specific P-splines baseline hazard function, given by $\log h^{\mbox{\tiny\text{T}}}_{0_k}(t)=\sum_{q=1}^{Q}\gamma^{\mbox{\tiny\text{T}}}_{0_{k_q}}\text{bs}^{\mbox{\tiny\text{T}}}_{k_q}\left(t\right)$, while $w^{\mbox{\tiny\text{T}}}_{k,i}(t)$ is the vector of observed (baseline or time-varying exogenous) explanatory variables, and $\boldsymbol{\gamma}^{\mbox{\tiny\text{T}}}_k$ is the corresponding vector of regression coefficients.

The $j$th longitudinal response influences the risk of failure of cause $k$ through $\mathcal{H}^{\mbox{\tiny\text{T}}}_{k,j,m}\left\{\eta_{j,i}(t)\right\}$. The association parameters $\alpha^{\mbox{\tiny\text{T}}}_{k,j,m}$ measure the strength of the association between each longitudinal outcome and the risk of the corresponding event. For a one-unit increase in $\mathcal{H}^{\mbox{\tiny\text{T}}}_{k,j,m}\left\{\eta_{j,i}(t)\right\}$, the HR for cause $k$ is $\exp(\alpha^{\mbox{\tiny\text{T}}}_{k,j,m})$. The longitudinal measurements and event times are assumed to be conditionally independent given $(\textbf b^\top_{1,i},\cdots,\textbf b^\top_{J,i})$.

The $k$th competing event is associated with the recurrent event process through a zero-mean Gaussian random variable $\upsilon^{\mbox{\tiny\text{T}}}_{k,i}$. We assume that the frailties $\upsilon^{\mbox{\tiny\text{T}}}_{k,i}$ and $\upsilon^{\mbox{\tiny\text{R}}}_i$ are proportional, $\upsilon^{\mbox{\tiny\text{T}}}_i=\upsilon^{\mbox{\tiny\text{R}}}_i\,\alpha^{\upsilon}_{k}$, reflecting the common underlying factors that affect their risk. The magnitude of the association between each pair of processes is quantified by $\alpha^\upsilon_k$, the log HR for a one-unit increase in the frailty term. We assume that correlations among different competing risks are driven by the shared frailty $\upsilon^{\mbox{\tiny\text{R}}}_i$. Conditional on $\upsilon^{\mbox{\tiny\text{R}}}_i$, the competing risks are independent of themselves and of the recurrent event times.

\subsection{Forms of association} \label{sec:meth:forms}
 
It has been recognized that the functional form used to link the longitudinal and event processes plays an important role in joint models~\citep{rizopoulos2014combining, mauff2017extension}. As discussed in Sections~\ref{sec:meth:rec} and \ref{sec:meth:term}, the hazards $h_i^{\mbox{\tiny\text{R}}}(t)$ and $h_i^{\mbox{\tiny\text{T}}}(t)$ of an event for patient $i$ at time $t$ are associated with the $j$th subject-specific marker trajectory through $\mathcal{H}^{\mbox{\tiny\text{R}}}_{j,m}\left\{\eta_{j,i}(t)\right\}$ and $\mathcal{H}^{\mbox{\tiny\text{T}}}_{k,j,m}\left\{\eta_{j,i}(t)\right\}$, respectively. Our model allows the specification of various forms of association between the longitudinal and time-to-event processes, such as underlying value, $\eta_{j,i}(t)$; slope, $\text{d}\eta_{j,i}(t)/\text{d}t$; standardized cumulative effect, $\frac{1}{t}\int_{0}^{t} \eta_{j,i}(s) \,\text{d} s$; and combinations of these regarding the same longitudinal outcome. Different forms can be assumed for each risk model. 

When a nonlinear link function $\mathcal{G}\left(\cdot\right)$ is applied to the mean of the longitudinal outcome in \eqref{eq:mth:meff}, it may be challenging to interpret the associations $\exp(\alpha^{\mbox{\tiny\text{T}}}_{k,j,m})$ and $\exp(\alpha^{\mbox{\tiny\text{R}}}_{j,m})$ in the linear predictor scale. In such situations, it is more convenient to transform the subject-specific linear predictor back to the outcome's original scale before applying the functional form of interest, that is, $\mathcal{H}_{j,m}\left\{\mu_{j,i}(t)\right\}=\mathcal{H}_{j,m}\left[\mathcal{G}^{-1}_j\left\{\eta_{j,i}(t)\right\}\right]$, where $\mathcal{G}^{-1}_j(\cdot)$ is the inverse link function. For example, when considering the logit link, we can use the expit function $\mathcal{G}^{-1}(x)=\mbox{expit}(x)=\exp(x)/\{1+\exp(x)\}$ so that the association parameters are interpretable in terms of the mean $\mu_{j,i}(t)$ of $y_{j,i}(t)$, and not in terms of $\mbox{logit}\left\{\mu_{j,i}(t)\right\}$. Supplementary~Table~S2 lists the functional forms that can be used in our model to link the longitudinal and time-to-event outcomes, along with the corresponding transformation functions.

\subsection{Inference and software} \label{sec:meth:inf}

Inference on the joint model parameters is carried out under the Bayesian framework. The corresponding posterior probability distribution does not have a closed form, so we resort to the Metropolis--Hastings algorithm with adaptive optimal scaling using the Robbins--Monro algorithm~\citep{garthwaite2016adaptive} to approximate it. Our C++ implementation of the posterior sampling algorithms allows fast model fitting despite its complexity and sample size, which have resulted in long computing times in previous analyses of the CFFPR~\citep{andrinopoulou2020multivariate}. The full and conditional posterior distributions, along with the prior specification, and additional details about the sampling heuristic, are available in Supplementary~Section~A.

We have made our model publicly available in the CRAN \textsf{R} package \texttt{JMbayes2}~\citep{jmbayes2}. In Supplementary~Section~B, we present an example of the use of the proposed joint model with \texttt{JMbayes2}. Our implementation allows the longitudinal processes to follow different distributions, such as the Student’s t, gamma, unit-Lindley, censored normal, binomial, Poisson, negative binomial, and beta-binomial distributions. Furthermore, the flexibility of our \texttt{JMbayes2} implementation allows users to fit simpler joint models that only consider the competing risks or the recurrent event processes.

\section{Simulation study} \label{sec:sim}
\subsection{Design} \label{sec:sim:dsgn}

The objective of our simulation study is twofold: to validate the proposed model and explore the bias introduced by model misspecification. We present two simulation scenarios, named A and B. Scenario~A is designed to validate the implementation of the model by demonstrating its ability to recover the parameters' true values. This scenario considers two longitudinal outcomes, two competing risks, and one recurrent process. The model structures for the data generation and fitting processes are identical. In Scenario~B, we examine the bias in the association parameter introduced by modeling a bounded outcome using a Gaussian distribution. This scenario involves a joint model with one longitudinal outcome and one terminal event. Two modeling strategies for the longitudinal submodel are considered: one using a beta distribution (the true model) and the other a Gaussian distribution (the misspecified model). The beta variant is used to assess the model under ideal conditions in which it is accurately specified, providing benchmark estimates for the Gaussian model. When considering the beta distribution, we include the longitudinal outcome in the hazards' linear predictors at its original scale, rather than the linear predictor scale, to ensure the comparability of association coefficients between the two models. 

Supplementary~Table~S3 provides the full definitions of the joint models employed for the data generation process and the corresponding models fitted to the generated data for both scenarios, and Supplementary~Table~S4 lists the parameter values considered. We replicate each scenario 100 times. Supplementary~Tables~S5~and~S6 detail the data generation process for each scenario, and Supplementary~Table~S7 summarizes the characteristics of the simulated datasets.

The joint models are fit using \verb+JMbayes2+ (v0.4.5). For each model, we use three Markov chains with 10,000 or 5,000 iterations per chain, discarding the first 7,500 and 2,500 iterations as a warm-up for Scenarios A and B, respectively. 
Details of the prior distributions assumed are available in the Supplementary~Table~S1. The convergence of the chains is assessed using the convergence diagnostic $\hat{R}$~\citep{gelman1992inference} aiming for values below 1.10, and by visual inspection of the posterior traceplots of randomly chosen datasets within each scenario. The code used to perform the simulation study is publicly available at \mbox{\url{https://github.com/pedromafonso/bounded-jm-simulation}}.

\subsection{Results} \label{sec:sim:res}

Table~\ref{tab:sim:est1} summarizes the simulation results, listing the bias and mean squared error values obtained. Supplementary~Figures~S1~and~S2 depict the distributions of estimated posterior means for both scenarios. In Scenario A, the estimates closely align with the true values, confirming the accuracy of the model. In Scenario B, the limitations of the Gaussian distribution become evident when dealing with inherently bounded longitudinal outcomes. Despite apparent convergence (see Supplementary~Figure~S3), the Gaussian model extrapolates the longitudinal model to values outside the response domain, introducing bias in the estimation of the target association (bias: -5.9; mean squared error [MSE]: 34.7) and, consequently, in the remaining independent variables present in the risk model. These findings underscore both the critical role of model selection and the suitability of the beta regression model for scenarios involving constrained response variables.

\begin{table}[!ht]
\begin{center}
\caption{Bias and mean squared error for the joint model estimates obtained under the two simulated scenarios for 100 simulated datasets. Scenario A: the joint model comprises one bounded and one unbounded longitudinal marker, two competing risks, and one recurrent event process; the fitted model is equal to the data generation model. Scenario B: the joint model comprises one bounded longitudinal marker and one terminal event; of the two fitted models, the one that models the bounded marker with a Gaussian distribution is different from the data generation model. Abbreviations: $\text{M}_1$, 1st longitudinal marker; $\text{M}_2$, 2nd longitudinal marker; MSE, mean squared error; PEx, pulmonary exacerbation; $\text{R}$, Recurrent event; $\text{T}_1$, 1st terminal event; $\text{T}_2$, 2nd terminal event.}
\label{tab:sim:est1}
\begin{tabular}{clrrrrrrrrrr} 
\toprule
& & \multicolumn{3}{c}{Scenario A} & & \multicolumn{6}{c}{Scenario B} \\
\cline{3-5} \cline{7-12} \\
& & & & & & & \multicolumn{2}{c}{Beta} & & \multicolumn{2}{c}{Gaussian} \\
\cline{8-9} \cline{11-12} \\
Submodel & Param. & True & Bias & MSE & & True & Bias & MSE & & Bias & MSE \\
\midrule
$\text{M}_1$ & & & & & & & & & & & \\
& $\beta_{1,0}$ & 2.00 & -0.001 & 0.000 & & 2.00 & -0.001 & 0.000 & & -1.235 & 1.526 \\
& $\beta_{1,t}$ & -1.50 & 0.001 & 0.000 & & -1.00 & 0.001 & 0.000 & & 0.881 & 0.777 \\
$\text{M}_2$ & & & & & & & & & & & \\
& $\beta_{2,0}$ & 0.80 & 0.000 & 0.000 & & -- & -- & -- & & -- & -- \\
& $\beta_{2,t}$ & -0.05 & 0.000 & 0.000 & & -- & -- & -- & & -- & -- \\
$\text{R}$ & & & & & & & & & & & \\
& $\gamma^{\mbox{\tiny\text{R}}}$ & 0.25 & -0.010 & 0.002 & & -- & -- & -- & & -- & -- \\
& $\alpha^{\mbox{\tiny\text{R}}}_1$ & -2.00 & -0.008 & 0.006 & & -- & -- & -- & & -- & -- \\
& $\alpha^{\mbox{\tiny\text{R}}}_2$ & -1.00 & -0.003 & 0.003 & & -- & -- & -- & & -- & -- \\
$\text{T}_1$ & & & & & & & & & & & \\
& $\gamma^{\mbox{\tiny\text{T}}}_{1}$ & 0.25 & -0.016 & 0.015 & & 0.25 & -0.004 & 0.006 & & -0.036 & 0.009 \\
& $\alpha^{\mbox{\tiny\text{T}}}_{1,1}$ & -2.00 & -0.079 & 0.378 & & -2.00 & -0.066 & 0.122 & & -5.870 & 34.696 \\
& $\alpha^{\mbox{\tiny\text{T}}}_{1,2}$ & -1.00 & -0.019 & 0.018 & & -- & -- & -- & & -- & -- \\
& $\alpha^\upsilon_1$ & 1.00 & 0.020 & 0.034 & & -- & -- & -- & & -- & -- \\
$\text{T}_2$ & & & & & & & & & & & \\
& $\gamma^{\mbox{\tiny\text{T}}}_{2}$ & 0.25 & -0.013 & 0.010 & & -- & -- & -- & & -- & -- \\
& $\alpha^{\mbox{\tiny\text{T}}}_{2,1}$ & -2.00 & -0.026 & 0.199 & & -- & -- & -- & & -- & -- \\
& $\alpha^{\mbox{\tiny\text{T}}}_{2,2}$ & -1.00 & -0.020 & 0.012 & & -- & -- & -- & & -- & -- \\
& $\alpha^\upsilon_2$ & 1.00 & -0.005 & 0.046 & & -- & -- & -- & & -- & -- \\
\bottomrule
\end{tabular}
\end{center}
\end{table}

\section{Application} \label{sec:app}

\subsection{The CFFPR dataset} \label{sec:app:data}

The CFFPR is one of the largest and most comprehensive databases of its kind, containing longitudinal clinical and demographic information on individuals living with CF in the US~\citep{knapp2016cystic}. Supplementary~Figure~S4 outlines the exclusion process applied to address data quality issues, such as missing data or data entry errors. The remaining data describe 23,543 individuals, who collectively contributed 1,315,586 observations between January 1, 2000, and December 31, 2017. The demographic, social, and clinical characteristics of the individuals analyzed are summarized in Supplementary~Table~S8. The baseline characteristics are ethnicity, genotype, birth cohort, and sex. The time-varying characteristics include pancreatic enzyme intake---implying pancreatic insufficiency---and environmental influences such as neighborhood material deprivation index (as defined by~\citealp{brokamp2019material}), percentage of green space\footnote[1]{Percentage of greenspace, impervious, and tree canopy areas within the Zone Improvement Plan Code Tabulation Area (ZCTA) derived from the National Land Cover Database~\citep{jin2019overall}.}, and moving-truck density. Previous research demonstrated that environmental and community characteristics, alongside clinical and demographic factors, are critical to comprehensively understand CF progression~\citep{gecili2023built, palipana2023social}.

\begin{figure}[!ht]
\centerline{\includegraphics[width=1.0\textwidth]{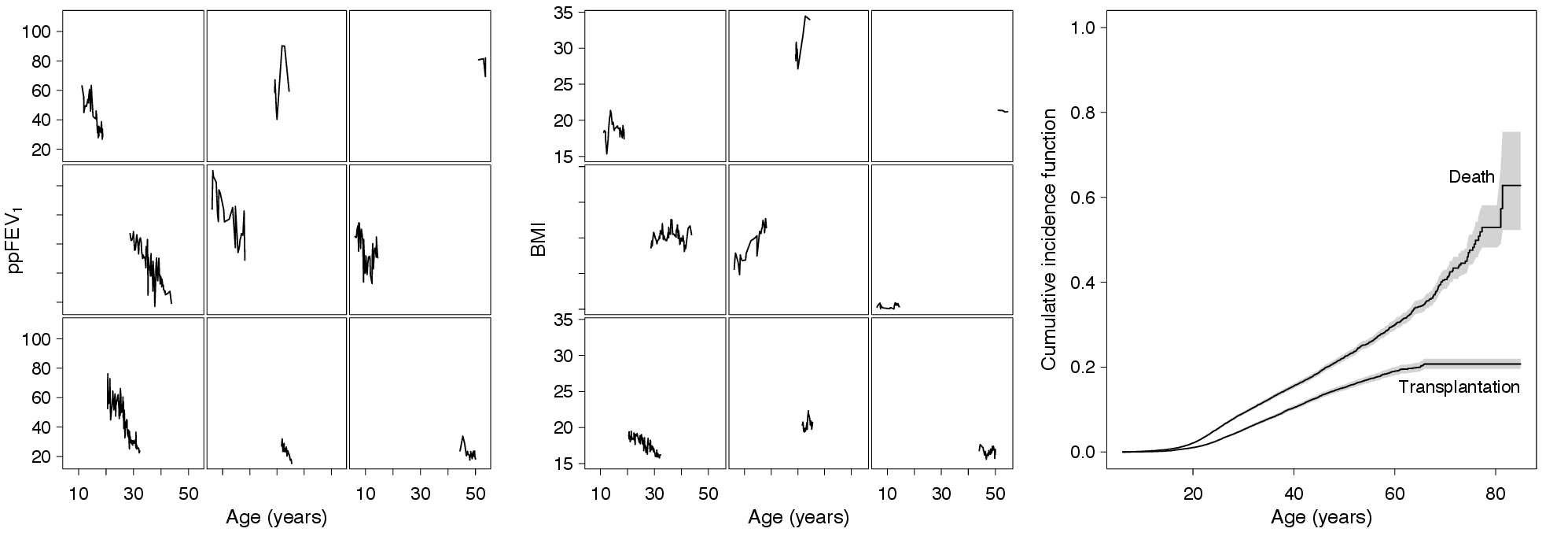}}
\caption{Longitudinal and survival outcomes of interest. Left:~ppFEV\textsubscript{1} measurements against age for nine randomly selected individuals. Center:~BMI measurements against age for the same individuals. Right:~Cumulative incidence functions for the competing events of death and lung transplantation, with associated 95\% confidence intervals.}
\label{fig:app:eda}
\end{figure}

BMI and ppFEV\textsubscript{1} are commonly measured in routine checkups and registered in the CFFPR\@. BMI is an important clinical marker used to assess the nutritional status of individuals with CF, who are at increased risk of malnutrition and poor growth due to impaired nutrient absorption, pancreatic insufficiency, and increased energy requirements. FEV\textsubscript{1} measures the maximum volume of air that a person can forcefully exhale in the first second of expiration after taking a deep breath. ppFEV\textsubscript{1} compares a patient's measured FEV\textsubscript{1} to the expected value for a person of the same age, sex, and height with normal lung function~\citep{stanojevic2015global}. We assume that ppFEV\textsubscript{1} ranges from 0\% to 150\%, with a value of 100\% meaning that the patient's FEV\textsubscript{1} is equal to the expected value for a healthy individual. While it is uncommon, there are instances in which the ppFEV\textsubscript{1} is reported as above 100\% owing to early intervention and treatment. Lower BMI and ppFEV\textsubscript{1} levels are associated with worse clinical outcomes~\citep{liou2001predictive}. The median numbers of ppFEV\textsubscript{1} and BMI measurements per individual are 47 (interquartile range [IQR] 27--69) and 48 (IQR 28--72), respectively, with corresponding median follow-up times per individual of 11.92 (IQR 6.97--16.76) and 11.72 (IQR 6.85--16.61) years. Figure~\ref{fig:app:eda} displays the ppFEV\textsubscript{1} (left~panel) and BMI (center~panel) evolution experienced by nine randomly selected individuals over time. The profiles exhibit different follow-up durations and diverse nonlinear trends.

The most common cause of death in cystic fibrosis patients is respiratory failure, often due to lung damage caused by chronic PEx. For individuals with end-stage lung disease, lung transplantation is a treatment option. Data acquired after lung transplantation were excluded. In this study, we treated death by respiratory failure and lung transplantation as competing events. However, formally, these events are semi-competing, as an individual can still die after receiving a double-lung transplant. Time-to-event data record the ages at which individuals experienced these events. During the follow-up period, 10.88\% of the individuals received a lung transplant,  17.97\% died from respiratory failure, and the remaining 71.15\% were right-censored. The median (IQR) ages at lung transplantation, death, and censoring were 28.52 (22.84--36.55), 26.57 (21.36--35.93), and 23.50 (17.07--32.15) years, respectively. The right~panel in Figure~\ref{fig:app:eda} shows the cumulative incidence functions for the competing risks of death and lung transplantation. We note that both of these events can cause nonignorable missing data in the measurements of ppFEV\textsubscript{1} and BMI.

A PEx is a sudden worsening of CF respiratory symptoms usually caused by an infection or inflammation in the airways~\citep{flume2009cystic}. In this study, we define the recurrent PEx event as an episode of care documented in the CFFPR with intravenous antibiotic use. If a new PEx episode is recorded during an ongoing exacerbation, it is treated as the same event. This implies the existence of non-risk periods during the episode of care that must be accounted for during the modeling process. The median number of PEx per individual is 7 (IQR 3--14), with a median interval between consecutive PEx of 0.34 (IQR 0.15--0.77) years.

\subsection{Analysis} \label{sec:app:anal}

We fitted the joint model described in Section~\ref{sec:meth}, considering two longitudinal outcomes ($J=2$), one recurrent event process, and two competing events ($K=2$). The longitudinal ppFEV\textsubscript{1} and BMI measurements are described using mixed-effects models assuming a beta and normal distribution, respectively. The formulations for these models are given as follows:
\begin{equation*}
\begin{aligned}
\mbox{logit}\left\{\text{ppFEV}^{\ast\ast}_{1\,i}(t)\right\}=&
\left(\beta_{1,0} + b_{1,0,i}\right) + \left(\beta_{1,\text{t}} + b_{1,t,i}\right)t + \beta_{1,\text{male}}\text{sex}_{\text{male},i}+\beta_{1,[93,98)}\text{YOB}_{[93, 98),i}+\\&+\beta_{1,\geq98}\text{YOB}_{\geq98,i}+\beta_{1,\text{htz}}\text{F508del}_{\text{htz},i}+\beta_{1,\text{oth}}\text{F508del}_{\text{oth},i}+\\
&+\beta_{1,\text{ethn}}\text{ethn}_{\text{hisp},i}+\beta_{1,\text{truck}}\text{truck}_i(t)+\beta_{1,\text{depr}}\text{depr}_i(t)
+\beta_{1,\text{pgrn}}\text{pgrn}_i(t),\\
\end{aligned}
\end{equation*}
and
\begin{equation*}
\begin{aligned}
\text{BMI}_i(t)=& \;\tildeL{\text{BMI}}_i(t) + \varepsilon_i(t) = 
\\ =& \left(\beta_{2,0} + b_{2,0,i}\right) + \sum_{q=1}^2\left(\beta_{2,q} + b_{2,q,i}\right)\text{ns}_{2,q}(t) + \beta_{2,\text{male}}\text{sex}_{\text{male},i}+\beta_{2,[93,98)}\text{YOB}_{[93, 98),i}+\\
&+\beta_{2,\geq98}\text{YOB}_{\geq98,i}+\beta_{2,\text{htz}}\text{F508del}_{\text{htz},i}+\beta_{2,\text{oth}}\text{F508del}_{\text{oth},i}+\beta_{2,\text{ethn}}\text{ethn}_{\text{hisp},i}+\\
&+\beta_{2,\text{depr}}\text{depr}_i(t)+\beta_{2,\text{enzy}}\text{enzy}_i(t)+\varepsilon_i(t),
\end{aligned}
\end{equation*}
for $t>0$, where $\left(b_{1,0,i}, b_{1,t,i}, b_{2,0,i}, b_{2,1,i},b_{2,2,i} b_{2,t^2,i}\right)^\top\sim \mathcal{N}\left(\textbf{0}, \textbf D\right)$, and $\varepsilon_i(t)\sim\mathcal{N}(0,\sigma^2_{\text{y}_2})$, with the two random variables assumed independent of each other. Here, $\tildeL{\text{BMI}}_i\nobreak(t)$ is the BMI response without error, and $\text{ppFEV}^{\ast\ast}_{1\,i}(t)$ is the ppFEV\textsubscript{1} response scaled to the interval $(0,1)$.\footnote[2]{A response restricted to a closed interval between known theoretical limits $a$ and $b$, so that $y\in\left[a, b\right]$, can be mapped to the interval $(0, 1)$ by transforming the observed value $y$ using $y^{\ast\ast}=\{y^\ast\times(N-1)+0.5\}/N$, where $y^\ast=(y-a)/(b-a)$ and $N$ is the sample size~\citep{smithson2006better}.}

For ppFEV\textsubscript{1}, we assume a linear average evolution over time, while for BMI, we assume a nonlinear evolution. More specifically, for BMI, we employ natural cubic splines with two degrees of freedom, denoted by $\text{ns}_{2,q}(t)$, $q=1,2$, with knots located at the 0\%, 50\% and 95\% percentiles of the observed follow-up times. 

The average ppFEV\textsubscript{1} and BMI responses are adjusted for baseline and time-varying individual characteristics including sex (male vs. female), $\text{sex}_{\text{male},i}$; birth cohort ($<93$, $[93, 98)$, or $\geq98$), $\text{YOB}_{<93,i}$ and $\text{YOB}_{[93,98),i}$; genotype (F508del homozygous, homozygous, or other/unknown), $\text{F508del}_{\text{htz},i}$ and $\text{F508del}_{\text{oth},i}$; ethnicity (hispanic vs. non-hispanic), $\text{ethn}_{\text{hisp},i}$; and neighborhood deprivation index, $\text{depr}_i(t)$. Additionally, the average ppFEV\textsubscript{1} is adjusted for the percentage of green space, $\text{pgrn}_i(t)$, and the annual average daily moving-truck density in the ZCTA, $\text{truck}_i(t)$, while the BMI response is adjusted for enzyme intake $\text{enzy}_i(t)$. The birth cohort variable aims to account for the evolution in CF care over the years, including approvals of new therapeutics. For the random effects structure, we assume a subject-specific random intercept and the same nonlinear effect of time as for the fixed effects.

We are interested in investigating how individual characteristics affect the risk of death separately from how they affect the risk of transplantation. Therefore, we postulate two cause-specific risk models, one for each of these competing events. The hazard functions for the clinical events of PEx, transplantation, and death are denoted by $h^{\mbox{\tiny\text{R}}}_{i}(t)$, $h^{\mbox{\tiny\text{T}}}_{1,i}(t)$, and $h^{\mbox{\tiny\text{T}}}_{2,i}(t)$, respectively, and are defined as follows
\begin{equation*}
h^{\mbox{\tiny\text{R}}}_i(t)= h^{\mbox{\tiny\text{R}}}_0(t-t_{0_{l,i}})\exp \Big[  \gamma^{\mbox{\tiny\text{R}}}_\text{PEx}\text{nPEx}_i(t) + 
\text{ppFEV}^{\ast\ast}_{1\,i}(t) \alpha^{\mbox{\tiny\text{R}}}_{1,1} +
\frac{1}{t}\int_{0}^{t} \tildeL{\text{BMI}}_i(s) \,\text{d} s\, \alpha^{\mbox{\tiny\text{R}}}_{2,1}  + \upsilon_i^{\mbox{\tiny\text{R}}}\Big],
\end{equation*}
\begin{equation*}
h^{\mbox{\tiny\text{T}}}_{1,i}(t)=h^{\mbox{\tiny\text{T}}}_{0_1}(t)\exp \Big[ \text{ppFEV}^{\ast\ast}_{1\,i}(t) \alpha^{\mbox{\tiny\text{T}}}_{1,1,1} + \frac{\text{d}\,\text{ppFEV}^{\ast\ast}_{1\,i}(t)}{\text{d}t} \alpha^{\mbox{\tiny\text{T}}}_{1,1,2}+
\frac{1}{t}\int_{0}^{t} \tildeL{\text{BMI}}_i(s) \,\text{d} s\, \alpha^{\mbox{\tiny\text{T}}}_{1,2,1} + \upsilon^{\mbox{\tiny\text{R}}}_i\,\alpha^\upsilon_1\Big],
\end{equation*}
and
\begin{equation*}
h^{\mbox{\tiny\text{T}}}_{2,i}(t)= h^{\mbox{\tiny\text{T}}}_{0_2}(t)\exp \Big[ \text{ppFEV}^{\ast\ast}_{1\,i}(t) \alpha^{\mbox{\tiny\text{T}}}_{2,1,1} + \frac{\text{d}\,\text{ppFEV}^{\ast\ast}_{1\,i}(t)}{\text{d}t} \alpha^{\mbox{\tiny\text{T}}}_{2,1,2}+  
\frac{1}{t}\int_{0}^{t} \tildeL{\text{BMI}}_i(s) \,\text{d} s\, \alpha^{\mbox{\tiny\text{T}}}_{2,2,1} + \upsilon^{\mbox{\tiny\text{R}}}_i\,\alpha^\upsilon_2\Big],
\end{equation*}
for $t>0$, where $\upsilon^{\mbox{\tiny\text{R}}}_i \sim \mathcal{N}\left(0, \sigma_\upsilon^2\right)$, $\upsilon^{\mbox{\tiny\text{R}}}_i \indep\left(b_{1,0,i}, b_{1,t,i}, b_{2,0,i}, b_{2,t,i}, b_{2,t^2,i}\right)$ and $\upsilon^{\mbox{\tiny\text{R}}}_i \indep\varepsilon_i(t)$. Changes in BMI over time occur relatively slowly, whereas ppFEV\textsubscript{1} can experience sudden declines. Therefore, guided by clinical insights, we include in the hazards' linear predictors the ppFEV\textsubscript{1}'s value, $\text{d}\,\text{ppFEV}^{\ast\ast}_1(t)/\text{d}t$, and rate of change, $\text{d}\,\text{ppFEV}^{\ast\ast}_1(t)/\text{d}t$, evaluated at its original scale---applying the $\mbox{expit}(\cdot)$ transformation to the linear predictor described in Section~\ref{sec:meth:long}---, and the standardized cumulative effect of BMI's underlying value, $\frac{1}{t}\int_{0}^{t} \tildeL{\text{BMI}}(s) \,\text{d} s$. In the PEx model we include the number of previous PEx events, $\text{nPEx}_i(t)$ and consider the gap timescale. Regarding the baseline hazards, we consider 10 quadratic P-spline basis functions defined over a grid of equally spaced knots over the domain of the observed event times. We consider second-order differences in the penalty matrices.

We generated three Markov chains in \verb+JMbayes2+ (v0.4.5) with 20,000 iterations each, of which 10,000 were discarded for warm-up. We use the package's default prior distributions (see Supplementary~Table~S1). The traceplots and the $\hat{R}$~\citep{gelman1992inference}, with $\hat{R}<1.10$, showed satisfactory convergence of the Markov chains.

\subsection{Results} \label{sec:app:res}

\begin{figure}
\centerline{\includegraphics[width=1.0\textwidth]{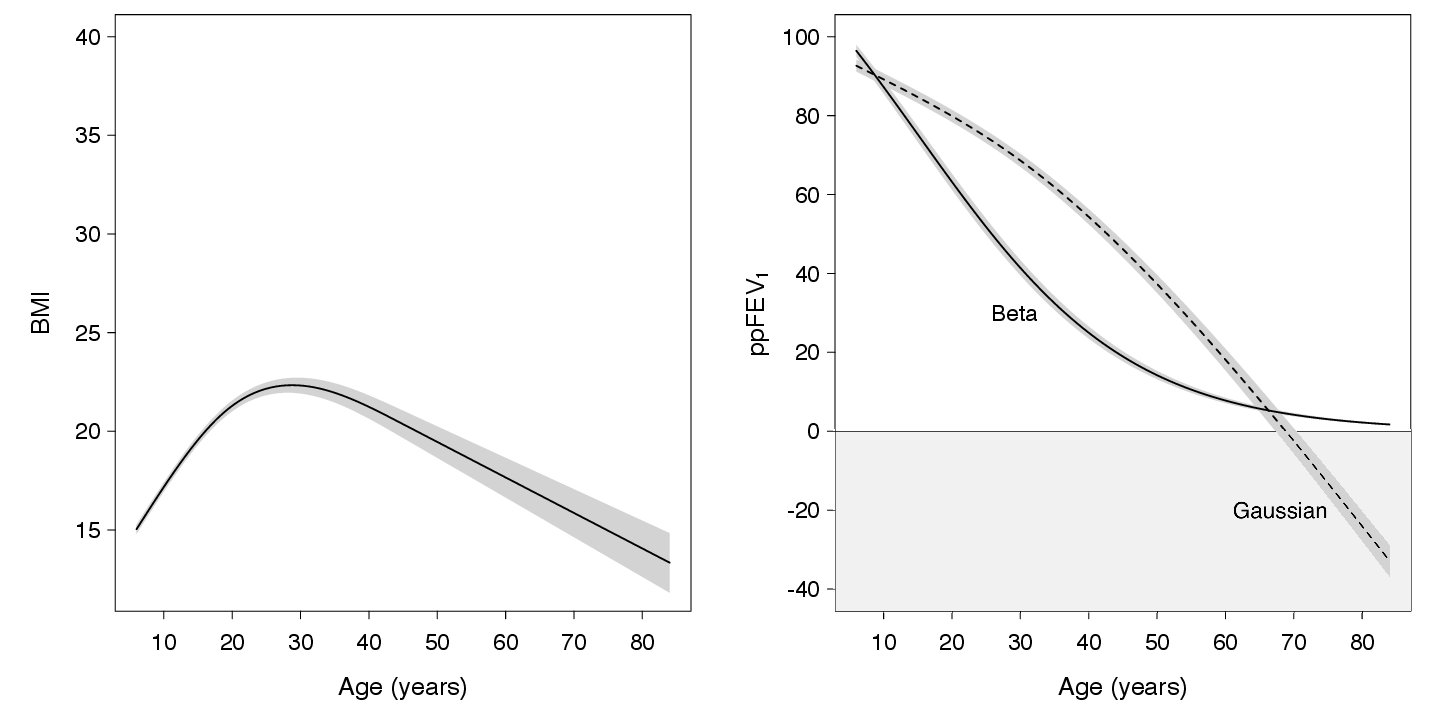}}
\caption{Left:~Estimated BMI evolution with age, with associated 95\% credible interval, for Hispanic females with CF, F508del homozygotes, who were born before 1993, did not take pancreatic enzymes, and lived in a community with deprivation index of 0.5. Right:~Estimated ppFEV\textsubscript{1} evolution with age, with associated 95\% credible interval, when assuming either a beta or Gaussian distribution for Hispanic females with CF, F508del homozygotes, who were born before 1993, 
and lived in a community with a deprivation index of 0.5, in which the percentage of green space is 50\%, and in which the moving-truck density is 0.18~\textmu truck-meters/m\textsuperscript{2}. For a Gaussian distribution, the model generates non-feasible negative values, despite incorporating flexible temporal evolution via natural cubic splines.}
\label{fig:app:effects}
\end{figure}

The effects plots in Figure~\ref{fig:app:effects} show the estimated evolution of BMI and ppFEV\textsubscript{1} with age. The results in the left~panel suggest an increase in BMI up to early adulthood, followed by a gradual decrease. The right~panel shows a period of rapid ppFEV\textsubscript{1} decline during childhood and adolescence, and a more gradual decline thereafter. When modeling ppFEV\textsubscript{1} with a Gaussian distribution and allowing for flexible temporal evolution, the resulting model produces non-feasible negative values (Figure~\ref{fig:app:effects}, right~panel).

The model parameter estimates are listed in Table~\ref{tab:app:est}. The risk of a PEx increases with the number of previous episodes. The results suggest that both ppFEV\textsubscript{1} and BMI are associated with the risks of experiencing PEx, transplantation, and death. A one-unit decrease in value and one-unit increase in the rate of ppFEV\textsubscript{1} decline increases the hazard of death by 11.58\% (95\%~CI~11.34--11.82) and 9.15\% (95\%~CI~7.51--10.83), respectively. A one-unit increase in the standardized cumulative effect of BMI increases the hazard of PEx by 7.06\% (95\%~CI~5.42--8.70). The incidence of PEx is positively associated with transplantation and death. Frailer individuals are at a higher risk of PEx and are more likely to receive a lung transplant or die. A one-standard-deviation increase in the frailty term increases the hazards of death
by 202.71\% (95\%~CI~187.69--219.03). In Supplementary~Section~D, the reader can find a detailed explanation of how these conclusions were derived from the association parameters estimates in Table~\ref{tab:app:est}. The estimates for the association between ppFEV\textsubscript{1} and the risk of transplantation are different from that between ppFEV\textsubscript{1} and death, illustrating the value of modeling both events individually, rather than as a composite endpoint.
\begin{table}[h]
\begin{center}
\caption{Posterior means and 95\% credible intervals for some of the joint model parameters fitted to the CFFPR dataset. Abbreviations: BMI, body mass index; CI, credible interval; HR, hazard ratio; PEx, pulmonary exacerbation; ppFEV\textsubscript{1}, percent predicted forced expiratory volume in one second.}
\label{tab:app:est}
\begin{tabular}{llrrr} 
\toprule
Model & Parameter/HR & Mean & \multicolumn{2}{c}{95\% CI}  \\
\midrule
ppFEV\textsubscript{1} & & & & \\
& $\beta_{1,0}$ & 0.591 & (0.554, & 0.629) \\
& $\beta_{1,t}$ & $-0.065$ & ($-0.065$, & $-0.064$) \\
& $\beta_{1,\text{male}}$ & 0.001 & ($-0.017$, & 0.018) \\
& $\beta_{1,\text{[93,98)}}$ & $-0.157$ & ($-0.180$, & $-0.133$) \\
& $\beta_{1,\geq98}$ & $-0.125$ & ($-0.147$, & $-0.103$) \\
& $\beta_{1,\text{htz}}$ & 0.019 & (0.001, & 0.038) \\
& $\beta_{1,\text{oth}}$ & $-0.024$ & ($-0.050$, & 0.002) \\
& $\beta_{1,\text{ethn}}$ & 0.223 & (0.191, & 0.256) \\
& $\beta_{1,\text{depr}}$ & $-0.003$ & ($-0.010$, & 0.005) \\
& $\beta_{1,\text{truck}}$ & $-4.44$e$^{-5}$ & ($-3.16$e$^{-4}$, & $3.34$e$^{-4}$) \\
& $\beta_{1,\text{pgrn}}$ & $-0.266$ & ($-0.519$, & $0.029$) \\
BMI & & & & \\
& $\beta_{2,0}$ & 15.053 & (14.858, & 15.244) \\
& $\beta_{2,\text{ns}_1}$ & 12.867 & (12.585, & 13.143) \\
& $\beta_{2,\text{ns}_2}$ & 1.881 & (1.424, & 2.330) \\
& $\beta_{2,\text{male}}$ & $-0.465$ & ($-0.548$, & $-0.378$) \\
& $\beta_{2,\text{[93,98)}}$ & 0.242 & (0.127, & 0.356) \\
& $\beta_{2,\geq98}$ & 0.633 & (0.523, & 0.743) \\
& $\beta_{2,\text{htz}}$ & 0.170 & (0.080, & 0.259) \\
& $\beta_{2,\text{oth}}$ & 0.269 & (0.140, & 0.398) \\
& $\beta_{2,\text{ethn}}$ & $-0.191$ & ($-0.348$, & $-0.032$) \\
& $\beta_{2,\text{depr}}$ & $-0.038$ & ($-0.101$, & $-0.021$) \\
& $\beta_{2,\text{enzy}}$ & 0.021 & (0.016, & 0.026) \\
Recurrent PEx & & & & \\
& $\exp(\gamma^{\mbox{\tiny\text{R}}}_\text{PEx})$ & 1.010 & (1.009, & 1.011) \\
& $\sigma_\upsilon$ & 0.835 & (0.822, & 0.849) \\
& $\exp(\alpha^{\mbox{\tiny\text{R}}}_{1,1}/150)$ & 0.962 & (0.961, & 0.962) \\
& $\exp(\alpha^{\mbox{\tiny\text{R}}}_{2,1})$ & 1.000 & (1.000, & 1.000) \\
Transplantation & & & & \\
& $\exp(\alpha^{\mbox{\tiny\text{T}}}_{1,1,1}/150)$ & 0.830 & (0.825, & 0.835) \\
& $\exp(\alpha^{\mbox{\tiny\text{T}}}_{1,1,2}/150)$ & 0.863 & (0.839, & 0.891) \\
& $\exp(\alpha^{\mbox{\tiny\text{T}}}_{1,2,1})$ & 1.060 & (1.044, & 1.076) \\
& $\exp(\alpha^\upsilon_1)$ & 1.203 & (1.122, & 1.287) \\
Death & & & & \\
& $\exp(\alpha^{\mbox{\tiny\text{T}}}_{2,1,1}/150)$ & 0.884 & (0.882, & 0.887) \\
& $\exp(\alpha^{\mbox{\tiny\text{T}}}_{2,1,2}/150)$ & 0.909 & (0.892, & 0.925) \\
& $\exp(\alpha^{\mbox{\tiny\text{T}}}_{2,2,1})$ & 1.071 & (1.054, & 1.087) \\
& $\exp(\alpha^\upsilon_2)$ & 1.326 & (1.266, & 1.389) \\
\bottomrule
\end{tabular}
\end{center}
\end{table}

\FloatBarrier

\section{Discussion} \label{sec:discuss}

Motivated by a clinical study on CF, we have developed the first Bayesian shared-parameter joint model that accommodates multiple continuous (possibly bounded) longitudinal markers, a recurrent event process, and multiple competing terminal events. Compared with previous frameworks, our comprehensive joint model enables more efficient use of all available information in scenarios with multiple markers and event times. In addition, by modeling a continuous and bounded longitudinal outcome using a beta distribution, we ensure that the longitudinal submodel predicts feasible values and provides meaningful insights into the association between the biomarker and the clinical event. This modeling framework can be particularly valuable for markers in pediatric populations expressed in percentiles or z-scores. The model is available in the \textsf{R} package \texttt{JMbayes2}~\citep{jmbayes2} and is flexible enough to handle a wide range of applications. 

The efficient implementation of the Markov chain Monte Carlo sampling algorithms in C++ ensures fast model fitting. Nonetheless, applying multivariate joint models to large datasets may require extended computing times. One can speed up model fitting by employing consensus Monte Carlo methods. Interested readers can find more details on how this approach can be implemented using \texttt{JMbayes2} in~\cite{afonso2023efficiently}.

It can be argued that all biomarkers are inherently bounded, as they signify measurable quantities within biological systems and are typically constrained by physiological limits. In the context of this study, BMI could be seen as inherently bounded like ppFEV\textsubscript{1}, making it a suitable candidate for modeling with a beta distribution. However, the normal distribution continues to be an effective approximation for BMI, as it will be for many other biomarkers, as the underlying distribution of the outcome lacks extreme skewness or heavy tails.

Although the proposed joint model exhibits great potential for advancing our understanding of complex disease dynamics, there remain opportunities for future research. We initially mapped the ppFEV\textsubscript{1} observations to the interval $\left[0, 1\right]$ and subsequently to the open interval $\left(0, 1\right)$ using the transformation proposed by~\cite{smithson2006better}. In future research, it may be worthwhile to explore the application of a zero-and-one inflated beta distribution to eliminate the need for the second transformation. Additionally, the derivation of individualized dynamic predictions~\citep{andrinopoulou2021reflection} represents an important research direction. Developing appropriate predictive assessment tools is also imperative for evaluating the model's performance and enabling its translation into clinical practice.

Our findings shed new light on the progression of CF, and we hope they will contribute to the effective management of PEx, reducing the frequency and severity of episodes. By making our model publicly available, we hope to assist applied statisticians and epidemiologists in performing joint analyses of longitudinal and time-to-event data in other complex settings.

\subsection*{Acknowledgments}
The authors would like to thank the Cystic Fibrosis Foundation for the use of CF Foundation Patient Registry data to conduct this study. Additionally, we would like to thank the patients, care providers, and clinic coordinators at CF Centers throughout the United States for their contributions to the CF Foundation Patient Registry.

\subsection*{Funding}
This work was supported by grants from the National Institutes of Health (R01 HL141286).

\subsection*{Data Availability Statement}
The data that support the findings of this study are available from the Cystic Fibrosis Foundation. Restrictions apply to the availability of these data, which were used under license for this study. Requests for data may be sent to datarequests@cff.org.

\bibliographystyle{biom}
\bibliography{references}

\begin{thebibliography}{}

\bibitem[\protect\citeauthoryear{Alsefri, Sudell, Garc{\'\i}a-Fi{\~n}ana, and Kolamunnage-Dona}{Alsefri et~al.}{2020}]{alsefri2020bayesian}
Alsefri, M., Sudell, M., Garc{\'\i}a-Fi{\~n}ana, M., and Kolamunnage-Dona, R. (2020).
\newblock Bayesian joint modelling of longitudinal and time to event data: A methodological review.
\newblock {\em BMC Medical Research Methodology} {\bf 20,} 1--17.

\bibitem[\protect\citeauthoryear{Andrinopoulou, Clancy, and Szczesniak}{Andrinopoulou et~al.}{2020}]{andrinopoulou2020multivariate}
Andrinopoulou, E.-R., Clancy, J.~P., and Szczesniak, R. (2020).
\newblock Multivariate joint modeling to identify markers of growth and lung function decline that predict cystic fibrosis pulmonary exacerbation onset.
\newblock {\em BMC Pulmonary Medicine} {\bf 20,} 1--11.

\bibitem[\protect\citeauthoryear{Andrinopoulou, Harhay, Ratcliffe, and Rizopoulos}{Andrinopoulou et~al.}{2021}]{andrinopoulou2021reflection}
Andrinopoulou, E.-R., Harhay, M.~O., Ratcliffe, S.~J., and Rizopoulos, D. (2021).
\newblock Reflection on modern methods: Dynamic prediction using joint models of longitudinal and time-to-event data.
\newblock {\em International Journal of Epidemiology} {\bf 50,} 1731--1743.

\bibitem[\protect\citeauthoryear{Andrinopoulou, Rizopoulos, Takkenberg, and Lesaffre}{Andrinopoulou et~al.}{2014}]{andrinopoulou2014joint}
Andrinopoulou, E.-R., Rizopoulos, D., Takkenberg, J.~J., and Lesaffre, E. (2014).
\newblock Joint modeling of two longitudinal outcomes and competing risk data.
\newblock {\em Statistics in Medicine} {\bf 33,} 3167--3178.

\bibitem[\protect\citeauthoryear{Bender, Augustin, and Blettner}{Bender et~al.}{2005}]{bender2005generating}
Bender, R., Augustin, T., and Blettner, M. (2005).
\newblock Generating survival times to simulate {Cox} proportional hazards models.
\newblock {\em Statistics in Medicine} {\bf 24,} 1713--1723.

\bibitem[\protect\citeauthoryear{Brokamp, Beck, Goyal, Ryan, Greenberg, and Hall}{Brokamp et~al.}{2019}]{brokamp2019material}
Brokamp, C., Beck, A.~F., Goyal, N.~K., Ryan, P., Greenberg, J.~M., and Hall, E.~S. (2019).
\newblock Material community deprivation and hospital utilization during the first year of life: An urban population--based cohort study.
\newblock {\em Annals of Epidemiology} {\bf 30,} 37--43.

\bibitem[\protect\citeauthoryear{Duchateau, Janssen, Kezic, and Fortpied}{Duchateau et~al.}{2003}]{duchateau2003evolution}
Duchateau, L., Janssen, P., Kezic, I., and Fortpied, C. (2003).
\newblock Evolution of recurrent asthma event rate over time in frailty models.
\newblock {\em Journal of the Royal Statistical Society: Series C (Applied Statistics)} {\bf 52,} 355--363.

\bibitem[\protect\citeauthoryear{Eilers and Marx}{Eilers and Marx}{1996}]{eilers1996flexible}
Eilers, P.~H. and Marx, B.~D. (1996).
\newblock Flexible smoothing with {B-splines} and penalties.
\newblock {\em Statistical Science} {\bf 11,} 89--121.

\bibitem[\protect\citeauthoryear{Elashoff, Li, and Li}{Elashoff et~al.}{2008}]{elashoff2008joint}
Elashoff, R.~M., Li, G., and Li, N. (2008).
\newblock A joint model for longitudinal measurements and survival data in the presence of multiple failure types.
\newblock {\em Biometrics} {\bf 64,} 762--771.

\bibitem[\protect\citeauthoryear{Farrell, Rosenstein, White, Accurso, Castellani, Cutting, Durie, LeGrys, Massie, Parad, et~al\mbox{.}}{Farrell et~al.}{2008}]{farrell2008guidelines}
Farrell, P.~M., Rosenstein, B.~J., White, T.~B., Accurso, F.~J., Castellani, C., Cutting, G.~R., Durie, P.~R., LeGrys, V.~A., Massie, J., Parad, R.~B., et~al. (2008).
\newblock Guidelines for diagnosis of cystic fibrosis in newborns through older adults: {Cystic Fibrosis Foundation} consensus report.
\newblock {\em The Journal of Pediatrics} {\bf 153,} S4--S14.

\bibitem[\protect\citeauthoryear{Faucett and Thomas}{Faucett and Thomas}{1996}]{faucett1996simultaneously}
Faucett, C.~L. and Thomas, D.~C. (1996).
\newblock Simultaneously modelling censored survival data and repeatedly measured covariates: A {Gibbs} sampling approach.
\newblock {\em Statistics in Medicine} {\bf 15,} 1663--1685.

\bibitem[\protect\citeauthoryear{Ferrari and Cribari-Neto}{Ferrari and Cribari-Neto}{2004}]{ferrari2004beta}
Ferrari, S. and Cribari-Neto, F. (2004).
\newblock Beta regression for modelling rates and proportions.
\newblock {\em Journal of Applied Statistics} {\bf 31,} 799--815.

\bibitem[\protect\citeauthoryear{Flume, Mogayzel~Jr, Robinson, Goss, Rosenblatt, Kuhn, Marshall, and {Clinical Practice Guidelines for Pulmonary Therapies Committee}}{Flume et~al.}{2009}]{flume2009cystic}
Flume, P.~A., Mogayzel~Jr, P.~J., Robinson, K.~A., Goss, C.~H., Rosenblatt, R.~L., Kuhn, R.~J., Marshall, B.~C., and {Clinical Practice Guidelines for Pulmonary Therapies Committee} (2009).
\newblock Cystic fibrosis pulmonary guidelines: Treatment of pulmonary exacerbations.
\newblock {\em American Journal of Respiratory and Critical Care Medicine} {\bf 180,} 802--808.

\bibitem[\protect\citeauthoryear{Garthwaite, Fan, and Sisson}{Garthwaite et~al.}{2016}]{garthwaite2016adaptive}
Garthwaite, P.~H., Fan, Y., and Sisson, S.~A. (2016).
\newblock Adaptive optimal scaling of metropolis--hastings algorithms using the robbins--monro process.
\newblock {\em Communications in Statistics-Theory and Methods} {\bf 45,} 5098--5111.

\bibitem[\protect\citeauthoryear{Gecili, Brokamp, Rasnick, Afonso, Andrinopoulou, Dexheimer, Clancy, Keogh, Ni, Palipana, et~al\mbox{.}}{Gecili et~al.}{2023}]{gecili2023built}
Gecili, E., Brokamp, C., Rasnick, E., Afonso, P.~M., Andrinopoulou, E.-R., Dexheimer, J.~W., Clancy, J.~P., Keogh, R.~H., Ni, Y., Palipana, A., et~al. (2023).
\newblock Built environment factors predictive of early rapid lung function decline in cystic fibrosis.
\newblock {\em Pediatric Pulmonology} .

\bibitem[\protect\citeauthoryear{Gelfand, Sahu, and Carlin}{Gelfand et~al.}{1995}]{gelfand1995efficient}
Gelfand, A.~E., Sahu, S.~K., and Carlin, B.~P. (1995).
\newblock Efficient parametrisations for normal linear mixed models.
\newblock {\em Biometrika} {\bf 82,} 479--488.

\bibitem[\protect\citeauthoryear{Gelman and Rubin}{Gelman and Rubin}{1992}]{gelman1992inference}
Gelman, A. and Rubin, D.~B. (1992).
\newblock Inference from iterative simulation using multiple sequences.
\newblock {\em Statistical Science} {\bf 7,} 457--472.

\bibitem[\protect\citeauthoryear{Gupta and Nadarajah}{Gupta and Nadarajah}{2004}]{gupta2004handbook}
Gupta, A.~K. and Nadarajah, S. (2004).
\newblock {\em Handbook of Beta Distribution and Its Applications}.
\newblock CRC Press.

\bibitem[\protect\citeauthoryear{Henderson, Diggle, and Dobson}{Henderson et~al.}{2000}]{henderson2000joint}
Henderson, R., Diggle, P., and Dobson, A. (2000).
\newblock Joint modelling of longitudinal measurements and event time data.
\newblock {\em Biostatistics} {\bf 1,} 465--480.

\bibitem[\protect\citeauthoryear{Hickey, Philipson, Jorgensen, and Kolamunnage-Dona}{Hickey et~al.}{2016}]{hickey2016joint}
Hickey, G.~L., Philipson, P., Jorgensen, A., and Kolamunnage-Dona, R. (2016).
\newblock Joint modelling of time-to-event and multivariate longitudinal outcomes: Recent developments and issues.
\newblock {\em BMC Medical Research Methodology} {\bf 16,} 1--15.

\bibitem[\protect\citeauthoryear{Hickey, Philipson, Jorgensen, and Kolamunnage-Dona}{Hickey et~al.}{2018}]{hickey2018joint}
Hickey, G.~L., Philipson, P., Jorgensen, A., and Kolamunnage-Dona, R. (2018).
\newblock Joint models of longitudinal and time-to-event data with more than one event time outcome: A review.
\newblock {\em The International Journal of Biostatistics} {\bf 14,}.

\bibitem[\protect\citeauthoryear{Jin, Homer, Yang, Danielson, Dewitz, Li, Zhu, Xian, and Howard}{Jin et~al.}{2019}]{jin2019overall}
Jin, S., Homer, C., Yang, L., Danielson, P., Dewitz, J., Li, C., Zhu, Z., Xian, G., and Howard, D. (2019).
\newblock Overall methodology design for the {United States} national land cover database 2016 products.
\newblock {\em Remote Sensing} {\bf 11,} 2971.

\bibitem[\protect\citeauthoryear{Kim, Zeng, Chambless, and Li}{Kim et~al.}{2012}]{kim2012joint}
Kim, S., Zeng, D., Chambless, L., and Li, Y. (2012).
\newblock Joint models of longitudinal data and recurrent events with informative terminal event.
\newblock {\em Statistics in Biosciences} {\bf 4,} 262--281.

\bibitem[\protect\citeauthoryear{Knapp, Fink, Goss, Sewall, Ostrenga, Dowd, Elbert, Petren, and Marshall}{Knapp et~al.}{2016}]{knapp2016cystic}
Knapp, E.~A., Fink, A.~K., Goss, C.~H., Sewall, A., Ostrenga, J., Dowd, C., Elbert, A., Petren, K.~M., and Marshall, B.~C. (2016).
\newblock The {Cystic Fibrosis Foundation Patient Registry}. {Design} and methods of a national observational disease registry.
\newblock {\em Annals of the American Thoracic Society} {\bf 13,} 1173--1179.

\bibitem[\protect\citeauthoryear{Kr{\'o}l, Ferrer, Pignon, Proust-Lima, Ducreux, Bouch{\'e}, Michiels, and Rondeau}{Kr{\'o}l et~al.}{2016}]{krol2016joint}
Kr{\'o}l, A., Ferrer, L., Pignon, J.-P., Proust-Lima, C., Ducreux, M., Bouch{\'e}, O., Michiels, S., and Rondeau, V. (2016).
\newblock Joint model for left-censored longitudinal data, recurrent events and terminal event: Predictive abilities of tumor burden for cancer evolution with application to the {FFCD} 2000--05 trial.
\newblock {\em Biometrics} {\bf 72,} 907--916.

\bibitem[\protect\citeauthoryear{Lang and Brezger}{Lang and Brezger}{2004}]{lang2004bayesian}
Lang, S. and Brezger, A. (2004).
\newblock Bayesian {P-splines}.
\newblock {\em Journal of Computational and Graphical Statistics} {\bf 13,} 183--212.

\bibitem[\protect\citeauthoryear{Liou, Adler, FitzSimmons, Cahill, Hibbs, and Marshall}{Liou et~al.}{2001}]{liou2001predictive}
Liou, T.~G., Adler, F.~R., FitzSimmons, S.~C., Cahill, B.~C., Hibbs, J.~R., and Marshall, B.~C. (2001).
\newblock Predictive 5-year survivorship model of cystic fibrosis.
\newblock {\em American Journal of Epidemiology} {\bf 153,} 345--352.

\bibitem[\protect\citeauthoryear{Liu and Huang}{Liu and Huang}{2009}]{liu2009joint}
Liu, L. and Huang, X. (2009).
\newblock Joint analysis of correlated repeated measures and recurrent events processes in the presence of death, with application to a study on acquired immune deficiency syndrome.
\newblock {\em Journal of the Royal Statistical Society: Series C (Applied Statistics)} {\bf 58,} 65--81.

\bibitem[\protect\citeauthoryear{Liu, Huang, and O'Quigley}{Liu et~al.}{2008}]{liu2008analysis}
Liu, L., Huang, X., and O'Quigley, J. (2008).
\newblock Analysis of longitudinal data in the presence of informative observational times and a dependent terminal event, with application to medical cost data.
\newblock {\em Biometrics} {\bf 64,} 950--958.

\bibitem[\protect\citeauthoryear{Mauff, Steyerberg, Nijpels, van~der Heijden, and Rizopoulos}{Mauff et~al.}{2017}]{mauff2017extension}
Mauff, K., Steyerberg, E.~W., Nijpels, G., van~der Heijden, A.~A., and Rizopoulos, D. (2017).
\newblock Extension of the association structure in joint models to include weighted cumulative effects.
\newblock {\em Statistics in Medicine} {\bf 36,} 3746--3759.

\bibitem[\protect\citeauthoryear{Miranda~Afonso, Rizopoulos, Palipana, Zhou, Brokamp, Szczesniak, and Andrinopoulou}{Miranda~Afonso et~al.}{2023}]{afonso2023efficiently}
Miranda~Afonso, P., Rizopoulos, D., Palipana, A.~K., Zhou, G.~C., Brokamp, C., Szczesniak, R.~D., and Andrinopoulou, E.-R. (2023).
\newblock Efficiently analyzing large patient registries with bayesian joint models for longitudinal and time-to-event data.
\newblock {\em arXiv preprint arXiv:2310.03351} .

\bibitem[\protect\citeauthoryear{Palipana, Vancil, Gecili, Rasnick, Ehrlich, Pestian, Andrinopoulou, Afonso, Keogh, Ni, et~al\mbox{.}}{Palipana et~al.}{2023}]{palipana2023social}
Palipana, A.~K., Vancil, A., Gecili, E., Rasnick, E., Ehrlich, D., Pestian, T., Andrinopoulou, E.-R., Afonso, P.~M., Keogh, R.~H., Ni, Y., et~al. (2023).
\newblock Social-environmental phenotypes of rapid cystic fibrosis lung disease progression in adolescents and young adults living in the united states.
\newblock {\em Environmental Advances} page 100449.

\bibitem[\protect\citeauthoryear{Papageorgiou, Mauff, Tomer, and Rizopoulos}{Papageorgiou et~al.}{2019}]{papageorgiou2019overview}
Papageorgiou, G., Mauff, K., Tomer, A., and Rizopoulos, D. (2019).
\newblock An overview of joint modeling of time-to-event and longitudinal outcomes.
\newblock {\em Annual Review of Statistics and Its Application} {\bf 6,} 223--240.

\bibitem[\protect\citeauthoryear{Rizopoulos}{Rizopoulos}{2012}]{rizopoulos2012joint}
Rizopoulos, D. (2012).
\newblock {\em Joint Models for Longitudinal and Time-to-Event Data: With Applications in R}.
\newblock CRC Press.

\bibitem[\protect\citeauthoryear{Rizopoulos and Ghosh}{Rizopoulos and Ghosh}{2011}]{rizopoulos2011bayesian}
Rizopoulos, D. and Ghosh, P. (2011).
\newblock A {Bayesian} semiparametric multivariate joint model for multiple longitudinal outcomes and a time-to-event.
\newblock {\em Statistics in Medicine} {\bf 30,} 1366--1380.

\bibitem[\protect\citeauthoryear{Rizopoulos, Hatfield, Carlin, and Takkenberg}{Rizopoulos et~al.}{2014}]{rizopoulos2014combining}
Rizopoulos, D., Hatfield, L.~A., Carlin, B.~P., and Takkenberg, J.~J. (2014).
\newblock Combining dynamic predictions from joint models for longitudinal and time-to-event data using {Bayesian} model averaging.
\newblock {\em Journal of the American Statistical Association} {\bf 109,} 1385--1397.

\bibitem[\protect\citeauthoryear{Rizopoulos, Papageorgiou, and {Miranda Afonso}}{Rizopoulos et~al.}{2022}]{jmbayes2}
Rizopoulos, D., Papageorgiou, G., and {Miranda Afonso}, P. (2022).
\newblock {\em JMbayes2: Extended Joint Models for Longitudinal and Time-to-Event Data}.
\newblock https://drizopoulos.github.io/JMbayes2/, https://github.com/drizopoulos/JMbayes2.

\bibitem[\protect\citeauthoryear{Smithson and Verkuilen}{Smithson and Verkuilen}{2006}]{smithson2006better}
Smithson, M. and Verkuilen, J. (2006).
\newblock A better lemon squeezer? maximum-likelihood regression with beta-distributed dependent variables.
\newblock {\em Psychological Methods} {\bf 11,} 54.

\bibitem[\protect\citeauthoryear{Sousa}{Sousa}{2011}]{sousa2011review}
Sousa, I. (2011).
\newblock A review on joint modelling of longitudinal measurements and time-to-event.
\newblock {\em Revstat Statistical Journal} {\bf 9,} 57--81.

\bibitem[\protect\citeauthoryear{Stanojevic, Bilton, McDonald, Stocks, Aurora, Prasad, Cole, and Davies}{Stanojevic et~al.}{2015}]{stanojevic2015global}
Stanojevic, S., Bilton, D., McDonald, A., Stocks, J., Aurora, P., Prasad, A., Cole, T.~J., and Davies, G. (2015).
\newblock {Global Lung Function Initiative} equations improve interpretation of {FEV1} decline among patients with cystic fibrosis.
\newblock {\em European Respiratory Journal} {\bf 46,} 262--264.

\bibitem[\protect\citeauthoryear{Szczesniak, Andrinopoulou, Su, Afonso, Burgel, Cromwell, Gecili, Ghulam, Goss, Mayer-Hamblett, et~al\mbox{.}}{Szczesniak et~al.}{2023}]{szczesniak2023lung}
Szczesniak, R., Andrinopoulou, E.-R., Su, W., Afonso, P.~M., Burgel, P.-R., Cromwell, E., Gecili, E., Ghulam, E., Goss, C.~H., Mayer-Hamblett, N., et~al. (2023).
\newblock Lung function decline in cystic fibrosis: Impact of data availability and modeling strategies on clinical interpretations.
\newblock {\em Annals of the American Thoracic Society} .

\bibitem[\protect\citeauthoryear{Tsiatis and Davidian}{Tsiatis and Davidian}{2004}]{tsiatis2004joint}
Tsiatis, A.~A. and Davidian, M. (2004).
\newblock Joint modeling of longitudinal and time-to-event data: An overview.
\newblock {\em Statistica Sinica} pages 809--834.

\bibitem[\protect\citeauthoryear{Williamson, Kolamunnage-Dona, Philipson, and Marson}{Williamson et~al.}{2008}]{williamson2008joint}
Williamson, P.~R., Kolamunnage-Dona, R., Philipson, P., and Marson, A.~G. (2008).
\newblock Joint modelling of longitudinal and competing risks data.
\newblock {\em Statistics in Medicine} {\bf 27,} 6426--6438.

\bibitem[\protect\citeauthoryear{Wulfsohn and Tsiatis}{Wulfsohn and Tsiatis}{1997}]{wulfsohn1997joint}
Wulfsohn, M.~S. and Tsiatis, A.~A. (1997).
\newblock A joint model for survival and longitudinal data measured with error.
\newblock {\em Biometrics} {\bf 53,} 330--339.

\end{thebibliography}

\newpage

\setcounter{table}{0} 
\renewcommand{\thetable}{S\arabic{table}} 
\setcounter{figure}{0} 
\renewcommand{\thefigure}{S\arabic{figure}} 

\section*{Supplementary material}

\appendix

\section{Posterior distribution} \label{sec:sup:est}

We denote the $j$th longitudinal marker measured at time $t$ for the $i$th individual by $Y_{j,i}(t),$ $i=1,\dots,n$, $j=1,\dots,J$. The longitudinal responses are collected for each subject at intermittent time points $\big\{t_{j,i,g},$ $j= 1,\ldots,J,$ $i = 1, \ldots,n,$ $ g = 1,\ldots,n_{j,i}\big\}$, where $n_{j,i}$ is the number of measurements of the longitudinal outcome $j$ for individual $i$, generating the vector of repeated measurements $\boldsymbol{Y}_{j,i}=(Y_{j,i,1},\ldots,Y_{j,i,n_{j,i}})^\top$, with $Y_{j,i,g}\equiv Y_{j,i}(t_{j,i,g})$. That is, $Y_{j,i,g}$ is the value of the $j$th longitudinal outcome for individual $i$ at time $t_{j,i,g}$. The number of measurements and the time points at which measurements are taken can differ between individuals, and a given individual can have different outcomes measured at different time points. Each individual may either experience one of the $K$ distinct competing terminal events or be right-censored during follow-up. Let $\text T_i$ denote the observed failure time for the $i$th individual, taken as $\text T_i=\min\left(\text T^\ast_{1,i}, \ldots, \text T^\ast_{K,i}, \text C_i\right)$, where $T^\ast_{k,i}$ is their true failure time for each event $k=1,\ldots,K$, and $\text C_i$ is the corresponding independent censoring time. The event indicator takes values $\delta^{\mbox{\tiny\text{T}}}_i\in\{0,1,\ldots,K\}$, with 0 corresponding to censoring and $1,\ldots,K$ to the competing terminal events. We assume that the missing values in the longitudinal measurements, aside from those caused by the $K$ events, are missing at random. Regarding the recurrent event process, let $\text R_{l,i}$ denote the time of the $l$th recurrent event experienced by the $i$th individual, $l=1,\ldots,L_i$, treated as $\text R_{l,i}=\min\left(\text R^\ast_{l,i},\text T_i\right)$, with $\text R^\ast_{l,i}$ being the $l$th true failure time. The event indicator $\delta^{\mbox{\tiny\text{R}}}_{l,i}$ is $1$ if $R^\ast_{l,i}<T_i$ and $0$ otherwise. Joint models assume a full joint distribution of the longitudinal and time-to-event processes $\left(\textbf Y_i, T_i, \textbf R_i\right)$, where $\boldsymbol{Y}_i=(\boldsymbol{Y}^\top_{1,i}, \ldots,\boldsymbol{Y}^\top_{J,i})^\top$ and $\textbf R_i=\left(R_{1,i},\ldots,R_{L_i,i}\right)^\top$. 

Let $\mathcal{D}_n=\big\{(\boldsymbol{Y}_i, T_i, \delta^{\mbox{\tiny\text{T}}}_i, \boldsymbol{R}_i, \boldsymbol{\delta}^{\mbox{\tiny\text{R}}}_i),$ $i=1,\ldots,n\big\}$ denote the observed information from a random sample of $n$ individuals of the target population, where $\boldsymbol{\delta}^{\mbox{\tiny\text{R}}}_i=(\delta^{\mbox{\tiny\text{R}}}_{1,i},\ldots,\delta^{\mbox{\tiny\text{R}}}_{L,i})^\top$. The unknown parameters $\boldsymbol{\theta}$, the subject-specific random effects $\textbf{b}=(\textbf{b}_1^\top,\ldots,\textbf{b}_n^\top)^\top$, and the frailty terms $\boldsymbol\upsilon=(\upsilon_1,\dots,\upsilon_n)^\top$ are estimated from the posterior distribution 
\begin{equation*}
\label{eq:meth:post}
p\left(\boldsymbol{\theta},\textbf{b},\boldsymbol\upsilon\mid\mathcal{D}_n\right)\propto p\left(\mathcal{D}_n\mid\boldsymbol{\theta},\textbf{b},\boldsymbol\upsilon\right)p\left(\boldsymbol{\theta},\textbf{b},\boldsymbol\upsilon\right),
\end{equation*}
where $p\left(\mathcal{D}_n\mid\boldsymbol{\theta},\textbf{b},\boldsymbol\upsilon\right)$ is the full likelihood of the model and $p\left(\boldsymbol{\theta},\textbf{b},\boldsymbol\upsilon\right)$ is the prior distribution. To evaluate the joint likelihood of the longitudinal and time-to-event outcome data, we assume that, given all observed covariates and the unobserved random effects, the longitudinal and survival processes are independent of each other, as are any given subject's longitudinal responses. Under this conditional independence assumption, the full likelihood can be written as
\begin{equation}\label{eq:meth:lh}
\begin{split}
p\left(\mathcal{D}_n\mid\boldsymbol{\theta},\textbf{b},\boldsymbol\upsilon\right)=&\prod_{i=1}^n\prod_{j=1}^J\prod_{g=1}^{n_{j,i}} p\left(Y_{j,i,g}\mid\boldsymbol{\theta}^{\mbox{\tiny\text{Y}}}_j,\boldsymbol{\theta}^{\mbox{\tiny\text{b}}}_j,\textbf{b}_{j,i}\right) p\left(T_i, \delta^{\mbox{\tiny\text{T}}}_i\mid\boldsymbol{\theta}^{\mbox{\tiny\text{T}}},\boldsymbol{\theta}^{\mbox{\tiny\text{Y}}},\boldsymbol{\theta}^{\mbox{\tiny\text{b}}},\theta^{\upsilon},\textbf{b}_i,\upsilon^{\mbox{\tiny\text{R}}}_i\right)\\
&\times\prod_{l=1}^{L_i} p\left(R_{l,i}, \delta^{\mbox{\tiny\text{R}}}_{l,i}\mid\boldsymbol{\theta}^{\mbox{\tiny\text{R}}},\boldsymbol{\theta}^{\mbox{\tiny\text{Y}}},\boldsymbol{\theta}^{\mbox{\tiny\text{b}}},\theta^{\upsilon},\textbf{b}_i,\upsilon^{\mbox{\tiny\text{R}}}_i\right)
p\left(\textbf{b}_i\mid\boldsymbol{\theta}^{\mbox{\tiny\text{b}}}\right) p\left(\upsilon^{\mbox{\tiny\text{R}}}_i\mid\theta^{\upsilon}\right), \\
\end{split}
\end{equation}
where $\boldsymbol{\theta}=\left(\boldsymbol{\theta}^{\mbox{\tiny\text{Y}}\top}, \boldsymbol{\theta}^{\mbox{\tiny\text{T}}\top}, \boldsymbol{\theta}^{\mbox{\tiny\text{R}}\top}, \boldsymbol{\theta}^{\mbox{\tiny\text{b}}\top}, \theta^{\upsilon}\right)^\top$ is the combined vector of unknown parameters. 
The longitudinal model parameters are denoted by $\boldsymbol{\theta}^{\mbox{\tiny\text{Y}}}=(\boldsymbol{\theta}^{\mbox{\tiny\text{Y}}\top}_1, \ldots, \boldsymbol{\theta}^{\mbox{\tiny\text{Y}}\top}_J)^\top$, with $\boldsymbol{\theta}^{\mbox{\tiny\text{Y}}}_j=(\boldsymbol{\beta}_j^\top,\sigma_{\text{y}_j})^\top$. The parameter vector of the competing risks models is represented by $\boldsymbol{\theta}^{\mbox{\tiny\text{T}}}=(\boldsymbol{\theta}^{\mbox{\tiny\text{T}}\top}_1, \ldots, \boldsymbol{\theta}^{\mbox{\tiny\text{T}}\top}_K)^\top$, with $\boldsymbol{\theta}^{\mbox{\tiny\text{T}}}_k=(\boldsymbol{\gamma}^{\mbox{\tiny\text{T}}\top}_k,\boldsymbol{\alpha}^{\mbox{\tiny\text{T}}\top}_k,\boldsymbol{\gamma}^{\mbox{\tiny\text{T}}\top}_{0_k},\alpha^{\upsilon}_k)^\top$. Finally, $\boldsymbol{\theta}^{\mbox{\tiny\text{R}}}=(\boldsymbol{\gamma}^{\mbox{\tiny\text{R}}\top},\boldsymbol{\alpha}^{\mbox{\tiny\text{R}}\top},\boldsymbol{\gamma}^{\mbox{\tiny\text{R}}\top}_0)^\top$ is the parameter vector of the recurrent time-to-event model, $\boldsymbol{\theta}^{\mbox{\tiny\text{b}}}\equiv\textbf D$ and $\theta_j^{\mbox{\tiny\text{b}}}\equiv\textbf D_j$ are the random-effects covariance matrices, and $\theta^{\upsilon}\equiv\sigma_\upsilon$ is the frailty standard deviation.

In~\eqref{eq:meth:lh}, the likelihood contribution $p\left(Y_{j,i,g}\mid\boldsymbol{\theta}^{\mbox{\tiny\text{Y}}}_j,\boldsymbol{\theta}^{\mbox{\tiny\text{b}}}_j,\textbf{b}_{j,i}\right)$ of the $g$th observation from the $j$th outcome of the $i$th individual is the probability mass or density function of the distribution considered. For a normal distribution, the contribution takes the form
\begin{equation*}
p\left(Y_{j,i,g}\mid\boldsymbol{\theta}^{\mbox{\tiny\text{Y}}}_j,\boldsymbol{\theta}^{\mbox{\tiny\text{b}}}_j,\textbf{b}_{j,i}\right)\propto\sigma_{y_j}^{-1/2}\exp\left\{-\frac{\left(Y_{j,i,g}-\textbf x_{j,i,g}^\top\boldsymbol\beta_j - \textbf z_{j,i,g}^\top\textbf b_{j,i}\right)^2}{2\sigma^2_{y_j}}\right\}.
\end{equation*}
When assuming a beta distribution, this is instead
\begin{equation*}
p\left(Y_{j,i,g}^\ast\mid\boldsymbol{\theta}^{\mbox{\tiny\text{Y}}}_j,\boldsymbol{\theta}^{\mbox{\tiny\text{b}}}_j,\textbf{b}_{j,i}\right)=\frac{\Gamma(\phi)}{\Gamma(\mu_{j,i,g}\,\phi)\Gamma\left((1-\mu_{j,i,g})\phi\right)}{Y^\ast_{j,i,g}}^{\mu_{j,i,g}\phi-1}(1-Y^\ast_{j,i,g})^{(1-\mu_{j,i,g})\phi-1},
\end{equation*}
where $\mu_{j,i,g}=\mathcal{G}^{-1}(\textbf x_{j,i,g}^\top\boldsymbol\beta_j + \textbf z_{j,i,g}^\top\textbf b_{j,i})$, $\Gamma(\cdot)$ denotes the gamma function, and $Y_{j,i,g}^\ast$ is the observed response $Y_{j,i,g}$ transformed to the standard unit interval, so that $Y^\ast_{j,i,g}\in(0, 1)$.

The $i$th likelihood contribution of the $K$ competing terminal events in~\eqref{eq:meth:lh} takes the form
\begin{equation}\label{eq:meth:liksurv}
\begin{split}
p\left(T_i, \delta^{\mbox{\tiny\text{T}}}_i\mid\boldsymbol{\theta}^{\mbox{\tiny\text{T}}},\boldsymbol{\theta}^{\mbox{\tiny\text{Y}}},\boldsymbol{\theta}^{\mbox{\tiny\text{b}}},\theta^{\upsilon},\textbf{b}_i,\upsilon^{\mbox{\tiny\text{R}}}_i\right)=&\prod_{k=1}^K\exp\Bigg[\gamma^{\mbox{\tiny\text{T}}}_{{0,k}_0}+\sum_{q=1}^{Q}\gamma^{\mbox{\tiny\text{T}}}_{{0,k}_q}\text{bs}^{\mbox{\tiny\text{T}}}_{k_q}\left(T_i\right) + \textbf w^{\mbox{\tiny\text{T}}\top}_{k,i}\left(T_i\right) \boldsymbol\gamma^{\mbox{\tiny\text{T}}}_k \\
&+ \sum_{j=1}^{J}\sum_{m=1}^{M_j} \mathcal{H}^{\mbox{\tiny\text{T}}}_{k,j,m}\left\{\eta_{j,i}\left(T_i\right)\right\}\alpha^{\mbox{\tiny\text{T}}}_{k,j,m} + \upsilon^{\mbox{\tiny\text{R}}}_i\,\alpha^{\upsilon}_{k}\Bigg]^{I\left(\delta^{\mbox{\tiny\text{T}}}_i=k\right)}\\
&\times\exp\Bigg(-\sum_{k=1}^K\exp(\upsilon^{\mbox{\tiny\text{R}}}_i\,\alpha^{\upsilon}_{k})\int_0^{t^{\mbox{\tiny\text{T}}}_{j,i}}\exp\bigg[\gamma^{\mbox{\tiny\text{T}}}_{{0,k}_0}+\\
&+\sum_{q=1}^{Q}\gamma^{\mbox{\tiny\text{T}}}_{{0,k}_q}\text{bs}^{\mbox{\tiny\text{T}}}_{k_q}\left(T_i\right) + \textbf w^{\mbox{\tiny\text{T}}\top}_{k,i}\left(T_i\right) \boldsymbol\gamma^{\mbox{\tiny\text{T}}}_k+ \\
&+ \sum_{j=1}^{J}\sum_{m=1}^{M_j} \mathcal{H}^{\mbox{\tiny\text{T}}}_{k,j,m}\left\{\eta_{j,i}\left(T_i\right)\right\}\alpha^{\mbox{\tiny\text{T}}}_{k,j,m}\bigg]\text{d}s\Bigg),
\end{split}
\end{equation}
where $I(\cdot)$ is the indicator function, $\text{bs}^{\mbox{\tiny\text{T}}}_{k_q}(t)$ is the P-splines' $q$th basis function of degree $d$, and $\gamma^{\mbox{\tiny\text{T}}}_{0_{k_q}}$ are the corresponding unknown coefficients for the baseline hazard. The likelihood contribution of the $l$th recurrent event experienced by the $i$th individual in~\eqref{eq:meth:lh} is given by 
\begin{equation}\label{eq:meth:likrec}
\begin{split}
p\left(R_{l,i}, \delta^{\mbox{\tiny\text{R}}}_{l,i}\mid\boldsymbol{\theta}^{\mbox{\tiny\text{R}}},\boldsymbol{\theta}^{\mbox{\tiny\text{Y}}},\boldsymbol{\theta}^{\mbox{\tiny\text{b}}},\theta^{\upsilon},\textbf{b}_i,\upsilon^{\mbox{\tiny\text{R}}}_i\right)=&
\exp\Bigg[\gamma^{\mbox{\tiny\text{R}}}_{0_0}+\sum_{q=1}^{Q}\gamma^{\mbox{\tiny\text{R}}}_{0_q}\text{bs}^{\mbox{\tiny\text{R}}}_q\left(R_{l,i}-t_{0_{l,i}}\right) + \textbf w^{\mbox{\tiny\text{R}}\top}_i\left(R_{l,i}\right) \boldsymbol\gamma^{\mbox{\tiny\text{R}}} \\
&+ \sum_{j=1}^{J}\sum_{m=1}^{M_j} \mathcal{H}^{\mbox{\tiny\text{R}}}_{j,m}\left\{\eta_{j,i}\left(R_{l,i}\right)\right\}\alpha^{\mbox{\tiny\text{R}}}_{j,m} + \upsilon^{\mbox{\tiny\text{R}}}_i\Bigg]^{\delta^{\mbox{\tiny\text{R}}}_{l,i}}\\
&\times\exp\Bigg(-\exp(\upsilon^{\mbox{\tiny\text{R}}}_i)\int_0^{t^{\mbox{\tiny\text{R}}}_{j,i}}\exp\bigg[\gamma^{\mbox{\tiny\text{R}}}_{0_0}+\\
&+\sum_{q=1}^{Q}\gamma^{\mbox{\tiny\text{R}}}_{0_q}\text{bs}^{\mbox{\tiny\text{R}}}_q\left(R_{l,i}-t_{0_{l,i}}\right)+\textbf w^{\mbox{\tiny\text{R}}\top}_i\left(R_{l,i}\right) \boldsymbol\gamma^{\mbox{\tiny\text{R}}}+ \\
&+\sum_{j=1}^{J}\sum_{m=1}^{M_j} \mathcal{H}^{\mbox{\tiny\text{R}}}_{j,m}\left\{\eta_{j,i}\left(R_{l,i}\right)\right\}\alpha^{\mbox{\tiny\text{R}}}_{j,m}\bigg]\text{d}s\Bigg),
\end{split}
\end{equation}
where $\text{bs}^{\mbox{\tiny\text{R}}}_q(t)$ is the P-splines' $q$th basis function of degree $d$, and $\gamma^{\mbox{\tiny\text{R}}}_{0_q}$ is the corresponding unknown coefficient for the baseline hazard.

The integrals in~\eqref{eq:meth:liksurv} and~\eqref{eq:meth:likrec} do not have analytical solutions. Thus we evaluate them using 
a 15-point Gauss--Kronrod quadrature rule, following \cite{rizopoulos2011bayesian}. The random effects $p\left(\textbf{b}_i\mid\boldsymbol{\theta}^{\mbox{\tiny\text{b}}}\right)$ and $p\left(\upsilon^{\mbox{\tiny\text{R}}}_i\mid\theta^{\upsilon}\right)$ contribute with the probability density functions of zero-mean multivariate and univariate Gaussian distributions, respectively.

The prior distributions considered for each parameter are listed in Table~\ref{tab:mth:prior}. We assume normal distributions for the fixed effects in both the longitudinal and survival submodels $\left\{\boldsymbol\beta_j, \boldsymbol\gamma^{\mbox{\tiny\text{T}}}_k\right\}$, and for the association coefficients $\left\{\alpha^{\mbox{\tiny\text{R}}}_{j,m}, \alpha^{\mbox{\tiny\text{T}}}_{k,j,m},\alpha^{\upsilon}_k\right\}$. We use gamma distributions for the standard deviations $\left\{\sigma_{y_j},\sigma_\upsilon\right\}$ of the frailty and error terms. For the covariance matrix $\textbf D$, we assume a Lewandowski--Kurowicka--Joe distribution. For the P-spline coefficients in the baseline hazards, we consider multivariate Gaussian distributions,
\begin{equation*}
\boldsymbol\gamma^{\mbox{\tiny\text{T}}}_{{0_k}}\mid\tau^{\mbox{\tiny\text{T}}}_k\sim N\left(\boldsymbol 0, \tau_k^{\mbox{\tiny\text{T}}} \textbf{M}^{\mbox{\tiny\text{T}}}_k\right),\;\tau^{\mbox{\tiny\text{T}}}_k\sim\text{Gam}\left(k^{\mbox{\tiny\text{T}}}_{{0_k}},\lambda^{\mbox{\tiny\text{T}}}_{{0_k}}\right),
\end{equation*}
\begin{equation*}
\boldsymbol\gamma^{\mbox{\tiny\text{R}}}_{0}\mid\tau^{\mbox{\tiny\text{R}}}\sim N\left(\boldsymbol 0, \tau^{\mbox{\tiny\text{R}}} \textbf{M}^{\mbox{\tiny\text{R}}}\right),\;\tau^{\mbox{\tiny\text{R}}}\sim\text{Gam}\left(k^{\mbox{\tiny\text{R}}}_0,\lambda^{\mbox{\tiny\text{R}}}_0\right),
\end{equation*}
where $\textbf{M}^{\mbox{\tiny\text{T}}}_k=\boldsymbol\Delta^{\mbox{\tiny\text{T}}\top}_{k,u}\boldsymbol\Delta^{\mbox{\tiny\text{T}}}_{k,u}+\textbf{I}\epsilon^{\mbox{\tiny\text{T}}}_k$ and $\textbf{M}^{\mbox{\tiny\text{R}}}=\boldsymbol\Delta^{\mbox{\tiny\text{R}}\top}_u\boldsymbol\Delta^{\mbox{\tiny\text{R}}}_u+\textbf{I}\epsilon^{\mbox{\tiny\text{R}}}$ are the penalty matrices such that $\boldsymbol\Delta^{\mbox{\tiny\text{T}}}_{k,u}$ and $\boldsymbol\Delta^{\mbox{\tiny\text{R}}}_u$ form $u$th-order differences of adjacent B-splines, and the terms $\textbf{I}\epsilon^{\mbox{\tiny\text{T}}}_k$ and $\textbf{I}\epsilon^{\mbox{\tiny\text{R}}}$ introduce a small ridge penalty. The smoothness of the splines is controlled by the gamma hyperpriors on $\tau^{\mbox{\tiny\text{T}}}_k$ and $\tau^{\mbox{\tiny\text{R}}}$. For more details on Bayesian P-splines, see the seminal work by  Lang~and~Brezge~\citep{lang2004bayesian}.

\begin{table}
\begin{center}
\caption{Prior distributions considered for each parameter in the proposed joint model. Abbreviations: LKJ, Lewandowski--Kurowicka--Joe.}
\label{tab:mth:prior}
\begin{tabular}{lll} 
\toprule
Outcome & Parameter & Prior  \\
\midrule
$j$th longitudinal marker & & \\
& $\boldsymbol\beta_j$ & $\text{Normal}\big(\boldsymbol\mu_{\beta_j},\boldsymbol\Sigma_{\beta_j}\big)$ \\
& $\textbf D$ & $\text{LKJ}\big(\eta\big)$  \\
& $\sigma_{y_j}$ & $\text{Gamma}\big(k_{y_j},\lambda_{y_j}\big)$ \\
Recurrent event & & \\
& $\boldsymbol\gamma^{\mbox{\tiny\text{R}}}_{0}\mid\tau^{\mbox{\tiny\text{R}}}$ & $ \text{Normal}\left(\boldsymbol 0, \tau^{\mbox{\tiny\text{R}}} \textbf{M}^{\mbox{\tiny\text{R}}}\right)$ \\
& $\tau^{\mbox{\tiny\text{R}}}$ & $\text{Gamma}\left(k^{\mbox{\tiny\text{R}}}_0,\lambda^{\mbox{\tiny\text{R}}}_0\right)$ \\
& $\gamma^{\mbox{\tiny\text{R}}}_{0_q}$ & $\text{Normal}\big(\mu_{\gamma_0^{\mbox{\tiny\text{R}}}},\sigma^2_{\gamma_0^{\mbox{\tiny\text{R}}}}\big)$ \\
& $\gamma^{\mbox{\tiny\text{R}}}$ & $\text{Normal}\big(\mu_{\gamma^{\mbox{\tiny\text{R}}}},\sigma^2_{\gamma^{\mbox{\tiny\text{R}}}}\big)$ \\
& $\alpha^{\mbox{\tiny\text{R}}}_{j,m}$ & $\text{Normal}\big(\mu_{\alpha_{j,m}^{\mbox{\tiny\text{R}}}},\sigma^2_{\alpha_{j,m}^{\mbox{\tiny\text{R}}}}\big)$ \\
& $\sigma_\upsilon$ & $\text{Gamma}\big(k_\upsilon,\lambda_\upsilon\big)$ \\
$k$th competing terminal event & & \\
& $\boldsymbol\gamma^{\mbox{\tiny\text{T}}}_{{0_k}}\mid\tau^{\mbox{\tiny\text{T}}}_k$ & $\text{Normal}\left(\boldsymbol 0, \tau_k^{\mbox{\tiny\text{T}}} \textbf{M}^{\mbox{\tiny\text{T}}}_k\right)$ \\
& $\tau^{\mbox{\tiny\text{T}}}_k$ & $\text{Gamma}\left(k^{\mbox{\tiny\text{T}}}_{{0_k}},\lambda^{\mbox{\tiny\text{T}}}_{{0_k}}\right)$ \\
& $\gamma^{\mbox{\tiny\text{T}}}_k$ & $\text{Normal}\big(\boldsymbol\mu_{\gamma_k^{\mbox{\tiny\text{T}}}},\boldsymbol\Sigma_{\gamma_k^{\mbox{\tiny\text{T}}}}\big)$ \\
& $\alpha^{\mbox{\tiny\text{T}}}_{k,j,m}$ & $\text{Normal}\big(\mu_{\alpha_{k,j,m}^{\upsilon}},\sigma^2_{\alpha_{k,j,m}^{\upsilon}}\big)$ \\
& $\alpha^{\upsilon}_k$ & $\text{Normal}\big(\mu_{\alpha_k^{\upsilon}},\sigma^2_{\alpha_k^{\upsilon}}\big)$ \\
\bottomrule
\end{tabular}
\end{center}
\end{table}
The conditional posterior distributions for the parameters $\boldsymbol{\theta}^{\mbox{\tiny\text{Y}}}_j=(\boldsymbol{\beta}_j,\sigma_{\text{y}_j})$ of the $j$th longitudinal outcome and the covariance matrix $\textbf D$ are
\begin{equation*}\label{eq:post:beta}
\begin{split}
p\left(\boldsymbol{\beta}_j\mid\mathcal{D}_n,
\textbf{b}, \boldsymbol\upsilon\right)\propto&
\prod_{i=1}^{n}
\prod_{g=1}^{n_{j,i}} p\left(Y_{j,i,g}\mid\boldsymbol{\theta}^{\mbox{\tiny\text{Y}}}_j,\boldsymbol{\theta}^{\mbox{\tiny\text{b}}}_j,\textbf{b}_{j,i}\right)
p\left(T_i, \delta^{\mbox{\tiny\text{T}}}_i\mid\boldsymbol{\theta}^{\mbox{\tiny\text{T}}},\boldsymbol{\theta}^{\mbox{\tiny\text{Y}}},\boldsymbol{\theta}^{\mbox{\tiny\text{b}}},\theta^{\upsilon},\textbf{b}_i,\upsilon^{\mbox{\tiny\text{R}}}_i\right)\\
&\times\prod_{l=1}^{L_i} p\left(R_{l,i}, \delta^{\mbox{\tiny\text{R}}}_{l,i}\mid\boldsymbol{\theta}^{\mbox{\tiny\text{R}}},\boldsymbol{\theta}^{\mbox{\tiny\text{Y}}},\boldsymbol{\theta}^{\mbox{\tiny\text{b}}},\theta^{\upsilon},\textbf{b}_i,\upsilon^{\mbox{\tiny\text{R}}}_i\right)
p\left(\textbf{b}_i\mid\boldsymbol{\theta}^{\mbox{\tiny\text{b}}}\right) p\left(\boldsymbol{\beta}_j\right),
\end{split}
\end{equation*}
\begin{equation*}\label{eq:post:sigmaY}
p\left(\sigma_{\text{y}_j}\mid\mathcal{D}_n,
\textbf{b}, \boldsymbol\upsilon\right)\propto 
\prod_{i=1}^{n}
\prod_{g=1}^{n_{j,i}} p\left(Y_{j,i,g}\mid\boldsymbol{\theta}^{\mbox{\tiny\text{Y}}}_j,\boldsymbol{\theta}^{\mbox{\tiny\text{b}}}_j,\textbf{b}_{j,i}\right)
p(\sigma_{\text{y}_j})
\end{equation*}
and
\begin{equation*}\label{eq:post:D}
p\left(\textbf D\mid\mathcal{D}_n,
\textbf{b}, \boldsymbol\upsilon\right)\propto
\prod_{i=1}^{n}
p\left(\textbf{b}_i\mid\boldsymbol{\theta}^{\mbox{\tiny\text{b}}}\right) 
p\left(\textbf D\right).
\end{equation*}

The conditional posterior distributions for the parameters $\boldsymbol{\theta}^{\mbox{\tiny\text{R}}}=(\boldsymbol{\gamma}^{\mbox{\tiny\text{R}}},\boldsymbol{\alpha}^{\mbox{\tiny\text{R}}},\boldsymbol{\gamma}^{\mbox{\tiny\text{R}}}_0)$ of the recurrent time-to-event outcome are
\begin{equation*}\label{eq:post:gammaR}
p\left(\boldsymbol{\gamma}^{\mbox{\tiny\text{R}}}\mid\mathcal{D}_n,
\textbf{b}, \boldsymbol\upsilon\right)\propto
\prod_{i=1}^{n}
\prod_{l=1}^{L_i} p\left(R_{l,i}, \delta^{\mbox{\tiny\text{R}}}_{l,i}\mid\boldsymbol{\theta}^{\mbox{\tiny\text{R}}},\boldsymbol{\theta}^{\mbox{\tiny\text{Y}}},\boldsymbol{\theta}^{\mbox{\tiny\text{b}}},\theta^{\upsilon},\textbf{b}_i,\upsilon^{\mbox{\tiny\text{R}}}_i\right)
p\left(\boldsymbol{\gamma}^{\mbox{\tiny\text{R}}}\right),
\end{equation*}
\begin{equation*}\label{eq:post:alphaR}
p\left(\boldsymbol{\alpha}^{\mbox{\tiny\text{R}}}\mid\mathcal{D}_n,
\textbf{b}, \boldsymbol\upsilon\right)\propto
\prod_{i=1}^{n}
\prod_{l=1}^{L_i} p\left(R_{l,i}, \delta^{\mbox{\tiny\text{R}}}_{l,i}\mid\boldsymbol{\theta}^{\mbox{\tiny\text{R}}},\boldsymbol{\theta}^{\mbox{\tiny\text{Y}}},\boldsymbol{\theta}^{\mbox{\tiny\text{b}}},\theta^{\upsilon},\textbf{b}_i,\upsilon^{\mbox{\tiny\text{R}}}_i\right)
p\left(\boldsymbol{\alpha}^{\mbox{\tiny\text{R}}}\right),
\end{equation*}
\begin{equation*}\label{eq:post:gammaR0}
p\left(\boldsymbol{\gamma}^{\mbox{\tiny\text{R}}}_0\mid\mathcal{D}_n,
\textbf{b},\boldsymbol\upsilon\right)\propto
\prod_{i=1}^{n}
\prod_{l=1}^{L_i} p\left(R_{l,i}, \delta^{\mbox{\tiny\text{R}}}_{l,i}\mid\boldsymbol{\theta}^{\mbox{\tiny\text{R}}},\boldsymbol{\theta}^{\mbox{\tiny\text{Y}}},\boldsymbol{\theta}^{\mbox{\tiny\text{b}}},\theta^{\upsilon},\textbf{b}_i,\upsilon^{\mbox{\tiny\text{R}}}_i\right)
p\left(\boldsymbol{\gamma}^{\mbox{\tiny\text{R}}}_0\right)
\end{equation*}
and
\begin{equation*}\label{eq:post:sigmaF}
p\left(\sigma_{\upsilon}\mid\mathcal{D}_n,
\textbf{b}, \boldsymbol\upsilon\right)\propto 
\prod_{i=1}^{n}
p\left(\upsilon^{\mbox{\tiny\text{R}}}_i\mid\theta^{\upsilon}\right)p\left(\sigma_{\upsilon}\right).
\end{equation*}

The conditional posterior distributions for the parameters $\boldsymbol{\theta}^{\mbox{\tiny\text{T}}}_k=(\boldsymbol{\gamma}^{\mbox{\tiny\text{T}}}_k,\boldsymbol{\alpha}^{\mbox{\tiny\text{T}}}_k,\boldsymbol{\gamma}^{\mbox{\tiny\text{T}}}_{0_k},\alpha^{\upsilon}_k)$ of the $k$th competing time-to-event outcome are
\begin{equation*}\label{eq:post:gammaT}
\begin{split}
p\left(\boldsymbol{\gamma}^{\mbox{\tiny\text{T}}}_k\mid\mathcal{D}_n, \textbf{b}, \boldsymbol\upsilon\right)\propto&
\prod_{i=1}^{n}
p\left(T_i, \delta^{\mbox{\tiny\text{T}}}_i\mid\boldsymbol{\theta}^{\mbox{\tiny\text{T}}},\boldsymbol{\theta}^{\mbox{\tiny\text{Y}}},\boldsymbol{\theta}^{\mbox{\tiny\text{b}}},\theta^{\upsilon},\textbf{b}_i,\upsilon^{\mbox{\tiny\text{R}}}_i\right)p\left(\boldsymbol{\gamma}^{\mbox{\tiny\text{T}}}_k\right),
\end{split}
\end{equation*}
\begin{equation*}\label{eq:post:alphaT}
\begin{split}
p\left(\boldsymbol{\alpha}^{\mbox{\tiny\text{T}}}_k\mid\mathcal{D}_n, \textbf{b}, \boldsymbol\upsilon\right)\propto&
\prod_{i=1}^{n}
p\left(T_i, \delta^{\mbox{\tiny\text{T}}}_i\mid\boldsymbol{\theta}^{\mbox{\tiny\text{T}}},\boldsymbol{\theta}^{\mbox{\tiny\text{Y}}},\boldsymbol{\theta}^{\mbox{\tiny\text{b}}},\theta^{\upsilon},\textbf{b}_i,\upsilon^{\mbox{\tiny\text{R}}}_i\right)p\left(\boldsymbol{\alpha}^{\mbox{\tiny\text{T}}}_k\right),
\end{split}
\end{equation*}
\begin{equation*}\label{eq:post:gammaT0}
\begin{split}
p\left(\boldsymbol{\gamma}^{\mbox{\tiny\text{T}}}_{0_k}\mid\mathcal{D}_n,
\textbf{b}, \boldsymbol\upsilon\right)\propto&
\prod_{i=1}^{n}
p\left(T_i, \delta^{\mbox{\tiny\text{T}}}_i\mid\boldsymbol{\theta}^{\mbox{\tiny\text{T}}},\boldsymbol{\theta}^{\mbox{\tiny\text{Y}}},\boldsymbol{\theta}^{\mbox{\tiny\text{b}}},\theta^{\upsilon},\textbf{b}_i,\upsilon^{\mbox{\tiny\text{R}}}_i\right)p\left(\boldsymbol{\gamma}^{\mbox{\tiny\text{T}}}_{0_k}\right),
\end{split}
\end{equation*}
and
\begin{equation*}\label{eq:post:alphaTF}
\begin{split}
p\left(\alpha^{\upsilon}_k\mid\mathcal{D}_n,
\textbf{b}, \boldsymbol\upsilon\right)\propto&
\prod_{i=1}^{n}
p\left(T_i, \delta^{\mbox{\tiny\text{T}}}_i\mid\boldsymbol{\theta}^{\mbox{\tiny\text{T}}},\boldsymbol{\theta}^{\mbox{\tiny\text{Y}}},\boldsymbol{\theta}^{\mbox{\tiny\text{b}}},\theta^{\upsilon},\textbf{b}_i,\upsilon^{\mbox{\tiny\text{R}}}_i\right)p\left(\alpha^{\upsilon}_k\right).
\end{split}
\end{equation*}

The conditional posterior distributions for the random effects $\textbf{b}_i$ and the frailty term $\upsilon^{\mbox{\tiny\text{R}}}_i$ are
\begin{equation*}\label{eq:post:re}
\begin{split}
p\left(\textbf{b}_i\mid\mathcal{D}_i,
\textbf{b}_i, \upsilon^{\mbox{\tiny\text{R}}}_i\right)\propto&
\prod_{j=1}^J\prod_{g=1}^{n_{j,i}} p\left(Y_{j,i,g}\mid\boldsymbol{\theta}^{\mbox{\tiny\text{Y}}}_j,\boldsymbol{\theta}^{\mbox{\tiny\text{b}}}_j,\textbf{b}_{j,i}\right)
p\left(T_i, \delta^{\mbox{\tiny\text{T}}}_i\mid\boldsymbol{\theta}^{\mbox{\tiny\text{T}}},\boldsymbol{\theta}^{\mbox{\tiny\text{Y}}},\boldsymbol{\theta}^{\mbox{\tiny\text{b}}},\theta^{\upsilon},\textbf{b}_i,\upsilon^{\mbox{\tiny\text{R}}}_i\right)\\
&\times\prod_{l=1}^{L_i} p\left(R_{l,i}, \delta^{\mbox{\tiny\text{R}}}_{l,i}\mid\boldsymbol{\theta}^{\mbox{\tiny\text{R}}},\boldsymbol{\theta}^{\mbox{\tiny\text{Y}}},\boldsymbol{\theta}^{\mbox{\tiny\text{b}}},\theta^{\upsilon},\textbf{b}_i,\upsilon^{\mbox{\tiny\text{R}}}_i\right)
p\left(\textbf{b}_i\mid\boldsymbol{\theta}^{\mbox{\tiny\text{b}}}\right)
\end{split}
\end{equation*}
and
\begin{equation*}\label{eq:post:frailty}
\begin{split}
p\left(\upsilon^{\mbox{\tiny\text{R}}}_i\mid\mathcal{D}_i,
\textbf{b}_i, \upsilon^{\mbox{\tiny\text{R}}}_i\right)\propto&
p\left(T_i, \delta^{\mbox{\tiny\text{T}}}_i\mid\boldsymbol{\theta}^{\mbox{\tiny\text{T}}},\boldsymbol{\theta}^{\mbox{\tiny\text{Y}}},\boldsymbol{\theta}^{\mbox{\tiny\text{b}}},\theta^{\upsilon},\textbf{b}_i,\upsilon^{\mbox{\tiny\text{R}}}_i\right)\\
&\times\prod_{l=1}^{L_i} p\left(R_{l,i}, \delta^{\mbox{\tiny\text{R}}}_{l,i}\mid\boldsymbol{\theta}^{\mbox{\tiny\text{R}}},\boldsymbol{\theta}^{\mbox{\tiny\text{Y}}},\boldsymbol{\theta}^{\mbox{\tiny\text{b}}},\theta^{\upsilon},\textbf{b}_i,\upsilon^{\mbox{\tiny\text{R}}}_i\right)
p\left(\upsilon^{\mbox{\tiny\text{R}}}_i\mid\theta^{\upsilon}\right).
\end{split}
\end{equation*}
We use hierarchical centering for the fixed effects of the longitudinal submodels~\citep{gelfand1995efficient}, and we standardize the covariates of the survival submodels to facilitate the convergence of the MCMC algorithms. Additionally, to speed up the sampling process, we perform parallel sampling of the random effects from different individuals and run the Markov chains in parallel on multiple processor cores.

\FloatBarrier
\newpage 

\section{An example with JMbayes2} \label{sec:sup:jmbayes2}

To fit the joint model, users are required to structure their data into two distinct datasets: one dedicated to capturing information related to competing risks and recurrent events (survival dataset), and another focused on longitudinal markers (longitudinal dataset). Below, we provide a subsample of simulated datasets intended as an illustration.

The survival dataset encompasses details pertaining to both competing risks and recurrent events as shown below. Each subject is represented by multiple rows, corresponding to the number of recurrent risk periods, plus one additional row for each competing event. The strata variable is essential to differentiate between various event processes.
\begin{verbatim}
  id tstart tstop status strata group
1  1   0.00  5.79      0      R     1
2  1   0.00  5.79      0    CR1     1
3  1   0.00  5.79      1    CR2     1
4  2   0.00  7.55      1      R     0
5  2   7.55  9.67      1      R     0
6  2   9.67 10.00      0      R     0
7  2   0.00 10.00      0    CR1     0
8  2   0.00 10.00      0    CR2     0
\end{verbatim}
The longitudinal dataset describes repeated measurements taken on the same subjects, organized in a long format. As shown below, each row corresponds to a single observation, and there might be multiple rows for each subject, representing different measurements over various time points.
\begin{verbatim}
  id time   y1   y2
1  1 0.00 0.89 0.77
2  1 0.26 0.84 0.76
4  1 2.09 0.29 0.69
5  2 0.00 0.93   NA
6  2 2.87 0.16 0.67
7  2 5.37 0.01 0.57
8  2 8.46   NA 0.46
9  2 8.85 0.02 0.46
\end{verbatim}
To adjust the joint model, users must first fit the mixed effects and proportional hazards submodels. Subsequently, these models are provided as arguments in the \texttt{jm()} function. Within the function call, users specify the preferred functional forms for the longitudinal outcomes in each relative-risk model, along with the chosen timescale. An illustrative example is presented below.

\newpage

\begin{verbatim}
# 1. Load the package
library(JMbayes2)
                   
# 2. Fit the longitudinal and survival submodels

# 2.1 Bounded longitudinal outcome (beta distribution)
beta_fit <- mixed_model(y2 ~ time treat, # fixed-effects formula
                        random = ~ time | id,  # random-effects formula 
                        family = beta.fam(), # distribution family
                        data = long) # longitudinal dataset

# 2.2 Unbounded longitudinal outcome (Gaussian distribution)
gaus_fit <- lme(y1 ~ time, # fixed-effects formula  
                random = ~ time | id,  # random-effects formula 
                data = long) # longitudinal dataset

# 2.3 Proportional hazards model
ph_fit <- coxph(Surv(tsart, tstop, status) ~ 
                    group : strata(strata), # model formula
                data = surv) # survival dataset

# 3. Fit the joint model that links the submodels
jm_fit <- jm(ph_fit, # survival submodel
             list(beta_fit, gaus_fit), # longitudinal submodels
             time_var = "time", # time variable in the longitudinal 
                                # submodels
             recurrent = "gap", # event timescale, or "calendar"
             functional_forms = ~ vexpit(value(y1)):strata # func-forms 
                                + value(y2)):strata)       # formula

summary(jm_fit)
\end{verbatim}

Further details about the package usage can be found on the dedicated website: \linebreak \url{https://drizopoulos.github.io/JMbayes2/}.

\begin{table}
\begin{center}
\caption{Functional forms available in the \textup{\textsf{R}} package \textup{\texttt{JMbayes2}} to link the longitudinal and time-to-event outcomes, and the associated transformation functions. \textsuperscript{\textdagger}: \texttt{velocity(}$\cdot$\texttt{)} can be used as an alias for \texttt{slope(}$\cdot$\texttt{)}.}
\label{tab:mth:forms}
\begin{tabular}{p{3.5cm}ll} 
\toprule
Functional form & Function & Argument  \\
\midrule
Underlying value & $\eta_{j,i}(t)$ & \verb+value(+$\cdot$\verb+)+\\
& $\log\left\{\eta_{j,i}(t)\right\}$ & \verb+vlog(value(+$\cdot$\verb+))+\\
& $\log_2\left\{\eta_{j,i}(t)\right\}$ & \verb+vlog2(value(+$\cdot$\verb+))+\\
& $\log_{10}\left\{\eta_{j,i}(t)\right\}$ & \verb+vlog10(value(+$\cdot$\verb+))+\\
& $\sqrt{\eta_{j,i}(t)}$ & \verb+vsqrt(value(+$\cdot$\verb+))+\\
& $\exp\left\{\eta_{j,i}(t)\right\}$ & \verb+vexp(value(+$\cdot$\verb+))+\\
& $\mbox{expit}\left\{\eta_{j,i}(t)\right\}$ & \verb+vexpit(value(+$\cdot$\verb+))+\\
& $a+b\,\eta_{j,i}(t)+c\,\eta_{j,i}^2(t)$ & \verb+poly2(value(+$\cdot$\verb+))+\\
& $a+b\,\eta_{j,i}(t)+c\,\eta_{j,i}^2(t)+d\,\eta_{j,i}^3(t)$ & \verb+poly3(value(+$\cdot$\verb+))+\\
& $a+b\,\eta_{j,i}(t)+c\,\eta_{j,i}^2(t)+d\,\eta_{j,i}^3(t)+e\,\eta_{j,i}^4(t)$ & \verb+poly4(value(+$\cdot$\verb+))+\\
& & \\
Slope & $\text{d}\,\eta_{j,i}(t)/\text{d}t$ &   \verb+slope(+$\cdot$\verb+)+\textsuperscript{\textdagger}\\
& $\left|\text{d}\,\eta_{j,i}(t)/\text{d}t\right|$ &   \verb+vabs(slope(+$\cdot$\verb+))+\\
& $\text{d}\exp\left\{\eta_{j,i}(t)\right\}/\text{d}t$ & \verb+Dexp(slope(+$\cdot$\verb+))+\\
& $\text{d}\mbox{expit}\left\{\eta_{j,i}(t)\right\}/\text{d}t$ & \verb+Dexpit(slope(+$\cdot$\verb+))+\\
& & \\
Acceleration & $\text{d}^2\eta_{j,i}(t)/\text{d}t^2$ &   \verb+acceleration(+$\cdot$\verb+)+\\
& & \\
Standardized cumulative effect & $\frac{1}{t}\int_{0}^{t} \eta_{j,i}(s) \,\text{d} s$ &  \verb+area(+$\cdot$\verb+)+\\
\bottomrule
\end{tabular}
\end{center}
\end{table}

\FloatBarrier
\newpage

\section{Simulation study} \label{sec:sup:sim}

\begin{figure}[H]
\centerline{\includegraphics[width=1.0\textwidth]{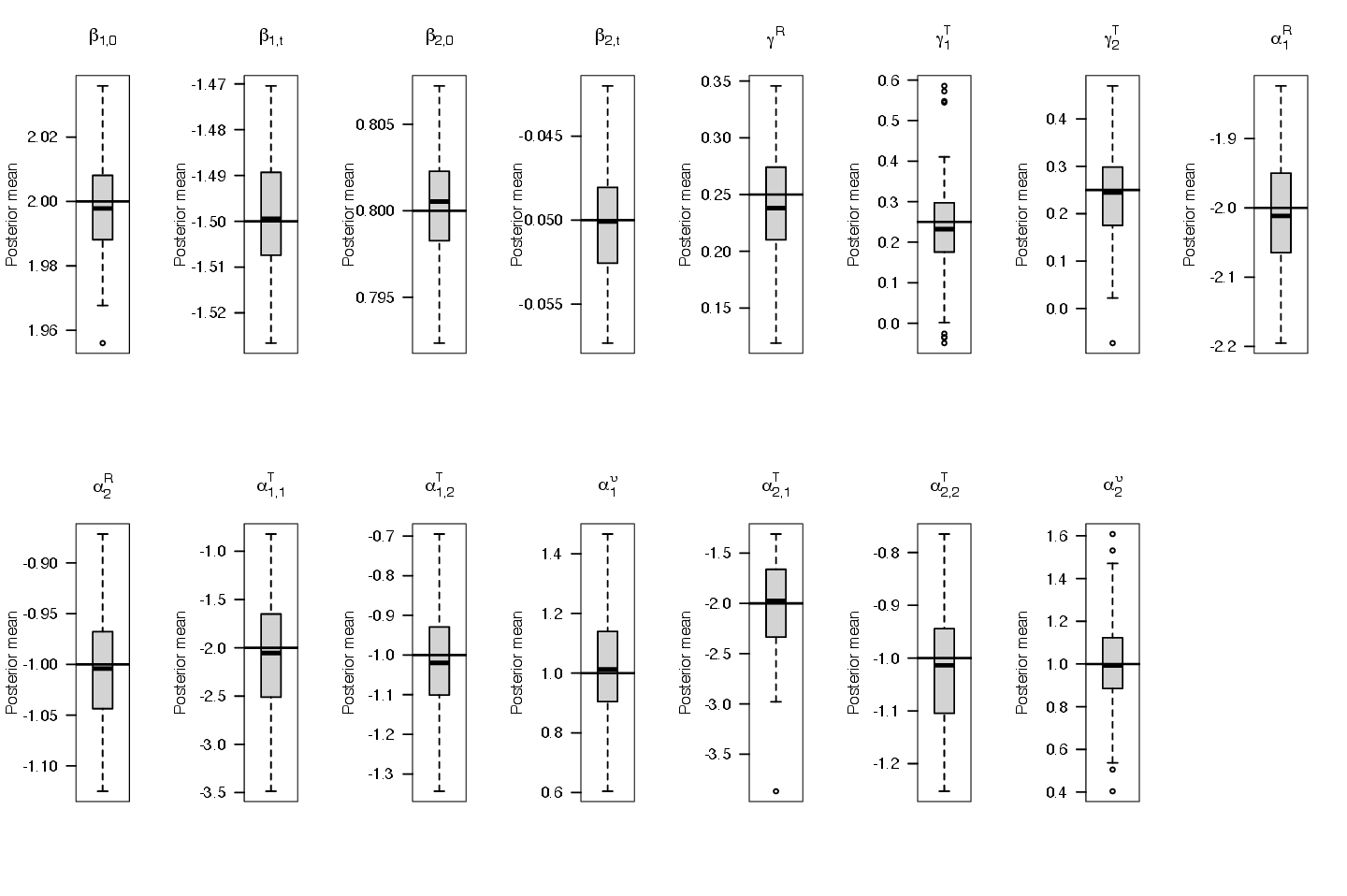}}
\caption{Estimated posterior means for joint model coefficients obtained in the simulation scenario A.}
\label{fig:sim:pmeanA}
\end{figure}

\begin{figure}
\centerline{\includegraphics[width=1.0\textwidth]{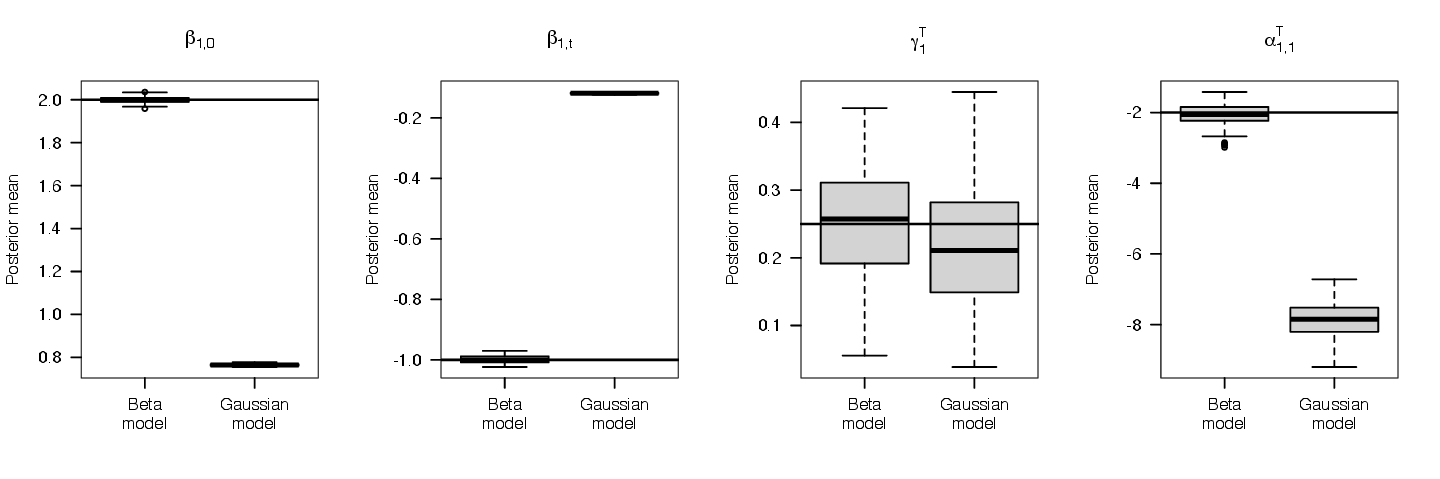}}
\caption{Estimated posterior means for joint model coefficients obtained in the simulation scenario B.}
\label{fig:sim:pmeanB}
\end{figure}

\begin{landscape}

\begin{table}
\begin{center}
\caption{Comparison of joint models considered under the simulation scenarios A and B. In scenario A, the fitted model is equal to the data generation model. In scenario B, the fitted model uses a Gaussian distribution to model the beta longitudinal outcome from the data generation model. We highlight the differences between the data and fitted models by enclosing varying elements within boxes in the model formulas. Abbreviations: $\text{M}_1$, 1st longitudinal marker; $\text{M}_2$, 2nd longitudinal marker; MSE, mean squared error; PEx, pulmonary exacerbation; $\text{R}$, recurrent event; $\text{T}_1$, 1st competing/terminal event; $\text{T}_2$, 2nd competing event.}\label{tab:sim:scn}
\begin{adjustbox}{max height=0.46\textheight}
\begin{tabular}{cllll} 
\toprule
& \multicolumn{1}{c}{Scenario A} & & \multicolumn{2}{c}{Scenario B} \\
\cline{2-2} \cline{4-5} \\
& \multicolumn{1}{c}{Data/Fit model} & & \multicolumn{1}{c}{Data model} & \multicolumn{1}{c}{Fit model} \\
\midrule
$\text{M}_1$ 
& $\begin{aligned}
&\mbox{logit}\left\{\mu_{1,i}(t)\right\}=\eta_{1,i}(t)=\\ 
&= (\beta_{1,0}+b_{1,0,i}) + (\beta_{1,t}+b_{1,t,i})t 
\end{aligned}$ 
& & $\begin{aligned}
&\boxed{\mbox{logit}\left\{\mu_{1,i}(t)\right\}}=\eta_{1,i}(t)=\\ 
&= (\beta_{1,0}+b_{1,0,i}) + (\beta_{1,t}+b_{1,t,i})t 
\end{aligned}$
& $\begin{aligned}
&\boxed{\mu_{1,i}(t)}=\eta_{1,i}(t)+\boxed{\varepsilon_{1,i}(t)}=\\ 
&= (\beta_{1,0}+b_{1,0,i}) + (\beta_{1,t}+b_{1,t,i})t +\boxed{\varepsilon_{1,i}(t)}
\end{aligned}$ \\
& & & \\
& $ (b_{1,0,i},b_{1,t,i})\sim\mathcal{N}\left(\textbf 0, \begin{bmatrix} \sigma^2_{1,0} & 0 \\ 0 & \sigma^2_{1, t} \end{bmatrix}\right)$ 
& & $ (b_{1,0,i},b_{1,t,i})\sim\mathcal{N}\left(\textbf 0, \begin{bmatrix} \sigma^2_{1,0} & 0 \\ 0 & \sigma^2_{1, t} \end{bmatrix}\right)$ 
& $ \begin{aligned}(b_{1,0,i},b_{1,t,i})&\sim\mathcal{N}\left(\textbf 0, \begin{bmatrix} \sigma^2_{1,0} & 0 \\ 0 & \sigma^2_{1, t} \end{bmatrix}\right) \\ 
\varepsilon_{1,i}(t)&\sim\mathcal{N}\left(0, \sigma^2_{1, t}\right) \end{aligned}$ \\
& & & \\
$\text{M}_2$
& $\begin{aligned}&\mu_{2,i}(t)=\eta_{2,i}(t)+\varepsilon_{2,i}(t)=\\ 
&= (\beta_{2,0}+b_{2,0,i}) + (\beta_{2,t}+b_{2,t,i})t +\varepsilon_{2,i}(t)
\end{aligned}$ 
& & -- & -- \\
& & & \\
& $ \begin{aligned}(b_{2,0,i},b_{2,t,i})&\sim\mathcal{N}\left(\textbf 0, \begin{bmatrix} \sigma^2_{2,0} & 0 \\ 0 & \sigma^2_{2, t} \end{bmatrix}\right) \\ 
\varepsilon_{2,i}(t)&\sim\mathcal{N}\left(0, \sigma^2_{2, t}\right) \end{aligned}$ & & & \\
& & & & \\
& & & \\
$\text{R}$ 
& $\begin{aligned}
&h^{\mbox{\tiny\text{R}}}_i(t)= h^{\mbox{\tiny\text{R}}}_0\left(t\right)\exp \Big[  w^{\mbox{\tiny\text{R}}}_i \gamma^{\mbox{\tiny\text{R}}} + \upsilon^{\mbox{\tiny\text{R}}}_i
\\&+ \mbox{expit}\left\{\eta_{1,i}(t)\right\}\alpha^{\mbox{\tiny\text{R}}}_1 + \eta_{2,i}(t)\alpha^{\mbox{\tiny\text{R}}}_2 \Big] 
\end{aligned}$ 
& & -- & -- \\
& & & \\
& & & \\
$\text{T}_1$ 
& $\begin{aligned}
&h^{\mbox{\tiny\text{T}}}_{1,i}(t)= h^{\mbox{\tiny\text{T}}}_{0_1}(t)\exp \Big[ w^{\mbox{\tiny\text{T}}}_{1,i}\gamma^{\mbox{\tiny\text{T}}}_1 + \upsilon^{\mbox{\tiny\text{R}}}_i\,\alpha^{\upsilon}_{1}
\\&+ \mbox{expit}\left\{\eta_{1,i}(t)\right\}\alpha^{\mbox{\tiny\text{T}}}_{1,1} + \eta_{2,i}(t)\alpha^{\mbox{\tiny\text{T}}}_{1,2} \Big]
\end{aligned}$ 
& & $\begin{aligned}
&h^{\mbox{\tiny\text{T}}}_{1,i}(t)= h^{\mbox{\tiny\text{T}}}_{0_1}(t)\exp \Big[ w^{\mbox{\tiny\text{T}}}_{1,i}\gamma^{\mbox{\tiny\text{T}}}_1
+ \boxed{\mbox{expit}\left\{\eta_{1,i}(t)\right\}}\alpha^{\mbox{\tiny\text{T}}}_{1,1} \Big]
\end{aligned}$ 
& $\begin{aligned}
&h^{\mbox{\tiny\text{T}}}_{1,i}(t)= h^{\mbox{\tiny\text{T}}}_{0_1}(t)\exp \Big[ w^{\mbox{\tiny\text{T}}}_{1,i}\gamma^{\mbox{\tiny\text{T}}}_1+
\boxed{\eta_{1,i}(t)}\alpha^{\mbox{\tiny\text{T}}}_{1,1} \Big]
\end{aligned}$ \\
& & & \\
& & & \\
$\text{T}_2$ 
& $\begin{aligned}
&h^{\mbox{\tiny\text{T}}}_{2,i}(t)= h^{\mbox{\tiny\text{T}}}_{0_2}(t)\exp \Big[w^{\mbox{\tiny\text{T}}}_{2,i} \gamma^{\mbox{\tiny\text{T}}}_2 + \upsilon^{\mbox{\tiny\text{R}}}_i\,\alpha^{\upsilon}_{2}
\\&+ \mbox{expit}\left\{\eta_{1,i}(t)\right\}\alpha^{\mbox{\tiny\text{T}}}_{2,1} + \eta_{2,i}(t)\alpha^{\mbox{\tiny\text{T}}}_{2,2} \Big]
\end{aligned}$ 
& & -- & -- \\
\bottomrule
\end{tabular}
\end{adjustbox}
\end{center}
\end{table}

\end{landscape}

\begin{figure}
\centerline{\includegraphics[width=1.0\textwidth]{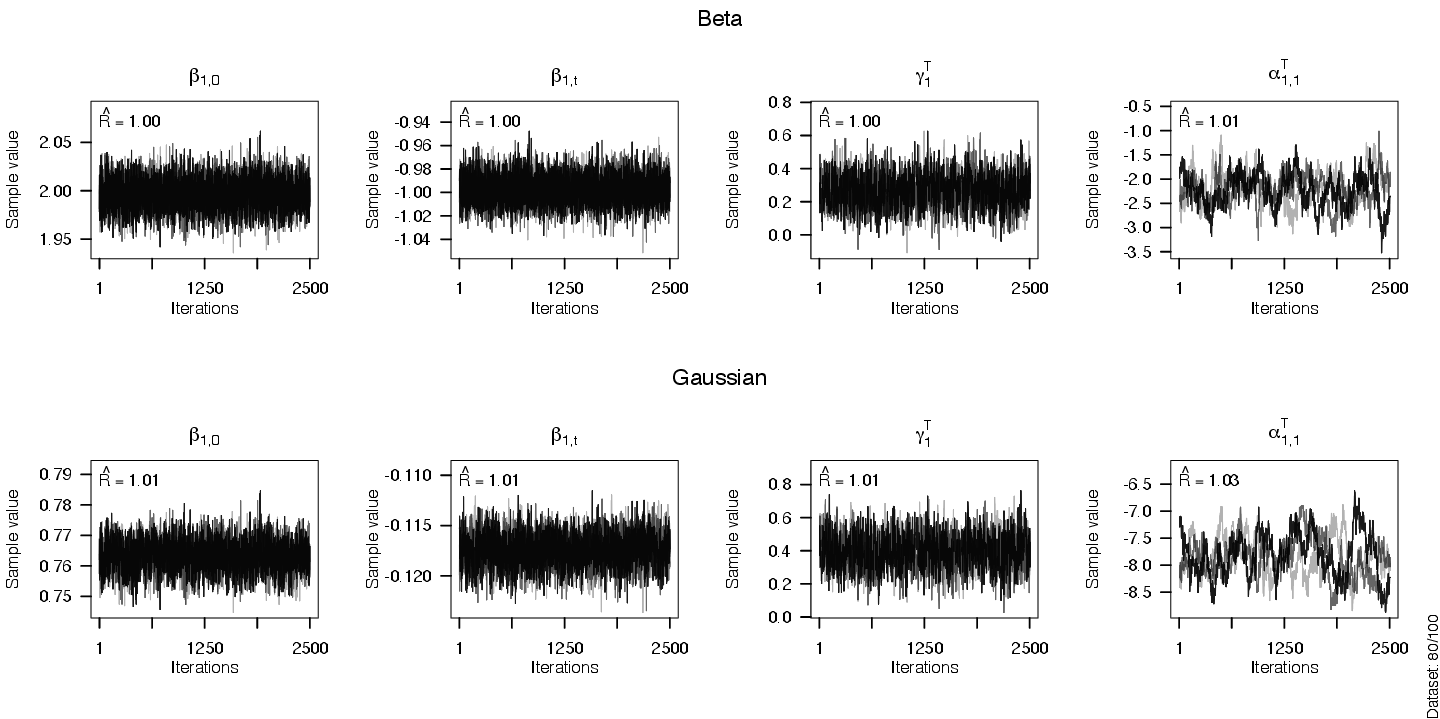}}
\caption{Traceplots for the joint model coefficients’ Markov chains after warm-up, for a randomly chosen dataset under scenario B. Top: Joint model with a beta submodel. Bottom: Joint model with a Gaussian submodel.}
\label{fig:sim:tcpB}
\end{figure}

\begin{table}
\begin{center}
\caption{Parameter values employed in the joint model for generating data in the simulation study. Abbreviations: $\text{M}_1$, 1st longitudinal marker; $\text{M}_2$, 2nd longitudinal marker; MSE, mean squared error; PEx, pulmonary exacerbation; $\text{R}$, recurrent event; $\text{T}_1$, 1st competing/terminal event; $\text{T}_2$, 2nd competing event.}\label{tab:sim:true}
\begin{tabular}{llrr} 
\toprule
& & \multicolumn{1}{c}{Scenario A} & \multicolumn{1}{c}{Scenario B} \\
\midrule
$\text{M}_1$ & & \\
& $\beta_{1,0}$ & 2.000 & 1.00 \\
& $\beta_{1,t}$ & -1.500 & -1.50 \\
& $\sigma_{1,0}$ & 0.250 & 0.25 \\
& $\sigma_{1,t}$ & 0.150 & 0.15 \\
& $\phi_{y}$ & $10^4$ & $10^4$ \\
$\text{M}_2$ & & \\
& $\beta_{2,0}$ & 0.800 & -- \\
& $\beta_{2,t}$ & -0.050 & -- \\
& $\sigma_{2,0}$ & 0.010 & -- \\
& $\sigma_{2,t}$ & 0.010 & -- \\
& $\sigma_{y}$ & 0.005 & -- \\
$\text{R}$ & & \\
& $h^{\mbox{\tiny\text{R}}}_{0_1}$ & 0.200 & -- \\
& $\gamma^{\mbox{\tiny\text{R}}}$ & 0.250 & -- \\
& $\alpha^{\mbox{\tiny\text{R}}}_1$ & -2.000 & -- \\
& $\alpha^{\mbox{\tiny\text{R}}}_2$ & -1.000 & -- \\
$\text{T}_1$ & & \\
& $h^{\mbox{\tiny\text{T}}}_{0_1}$ & 0.200 & 0.10 \\
& $\gamma^{\mbox{\tiny\text{T}}}_1$ & 0.250 & 0.25 \\
& $\alpha^{\mbox{\tiny\text{T}}}_{1,1}$ & -2.000 & -2.00 \\
& $\alpha^{\mbox{\tiny\text{T}}}_{1,2}$ & -1.000 & -- \\
& $\alpha^{\upsilon}_1$ & 1.000 & -- \\
$\text{T}_2$ & & \\
& $h^{\mbox{\tiny\text{T}}}_{0_2}$ & 2.000 & -- \\
& $\gamma^{\mbox{\tiny\text{T}}}_2$ & 0.250 & -- \\
& $\alpha^{\mbox{\tiny\text{T}}}_{2,1}$ & -2.000 & -- \\
& $\alpha^{\mbox{\tiny\text{T}}}_{2,2}$ & -1.000 & -- \\
& $\alpha^{\upsilon}_2$ & 1.000 & -- \\
\bottomrule
\end{tabular}
\end{center}
\end{table}

\begin{table}
\caption{Outline of the data generation process for scenario A.}
\label{tab:sim:jmA}
\centering
\begin{adjustbox}{max width=\textwidth}
\begin{tabular}{lp{16.3cm}}
\toprule
\multicolumn{2}{l}{Longitudinal outcome (1/2):} \\
1: & Generate $n=1000$ random samples from $\mathcal{N}\left(\boldsymbol 0 ,\boldsymbol \Sigma^{-1}\right)$ for the individual-specific random effects, $\boldsymbol b_i = (\underset{n\times2}{\boldsymbol b_{1,i}^\top}, \underset{n\times2}{\boldsymbol b_{2,i}^\top})$: $\underset{n\times4}{\boldsymbol b}$. \\
2: & Generate $(n\times (n_i-1))$ random samples from $\mathcal{U}\left(0,10\right)$ for the individual visiting times and add the time $0$, $\boldsymbol t_i$: $\underset{(n\cdot n_i)\times1}{\boldsymbol t}$.\\
3: & Generate the $(n\times2)$ vectors of $n_i$ individual underlying longitudinal responses, $\boldsymbol \mu_{1, i}$ and $\boldsymbol \mu_{2, i}$: $\boldsymbol \mu_{i,1}=\mbox{expit}(\boldsymbol \eta_{1,i})$ and $\boldsymbol \mu_{i,2}=\boldsymbol \eta_{2,i}$, where $\underset{n_i\times1}{\strut\boldsymbol \eta_{j,i}}= \underset{ n_i\times2}{\strut\begin{bmatrix} \boldsymbol 1 & \boldsymbol t_i \end{bmatrix}}\underset{2\times1}{\strut\boldsymbol\beta_j} + \underset{n_i\times2}{\begin{bmatrix} \boldsymbol 1 & \boldsymbol t_i \end{bmatrix}}\underset{2\times1}{\strut\boldsymbol b_i}$, $j=1,2$. \\
4: & Generate $(n\times n_i)$ random samples from
$\text{Beta}\left(p,q\right)$, where $p=\phi \times \mu_{1,i}(t)$ and $q=\phi \times \{1-\mu_{1,i}(t)\}$, for the observed beta longitudinal responses: $\underset{(n\cdot n_i)\times1}{\boldsymbol y_1}$.\\
5: & Generate $(n\times n_i)$ random samples from
$\mathcal{N}\left(0,\sigma^2_y\right)$ for the observation measurement error, $\varepsilon_{2,i}(t)$: $\underset{(n\cdot n_i)\times1}{\boldsymbol \varepsilon_2}$.\\
6: & Obtain the observed Gaussian longitudinal response by summing the vectors $\boldsymbol\eta_2$  and $\boldsymbol\varepsilon_2$: $\underset{(n\cdot n_i)\times1}{\boldsymbol y_2}$.\\
\midrule
\multicolumn{2}{l}{Survival outcome:} \\
7: & Generate $n$ random samples from
$\text{Bern}\left(0.5\right)$ for the individual's group, $w_i$: $\underset{n\times1}{\boldsymbol{w}}$.\\
8: & Generate $(n\times2)$ random samples from $\mathcal{U}\left(0, 1\right)$, $u^{\mbox{\tiny\text{T}}}_{1,i}$ and $u^{\mbox{\tiny\text{T}}}_{2,i}$: $\underset{n\times1}{\boldsymbol u^{\mbox{\tiny\text{T}}}_1}$ and $\underset{n\times1}{\boldsymbol u^{\mbox{\tiny\text{T}}}_2}$. \\
9: & Define $H^{\mbox{\tiny\text{T}}}_{j,i}(t)=\int_{0}^{t} h^{\mbox{\tiny\text{T}}}_{j,i}(t) \,ds$, where $h^{\mbox{\tiny\text{T}}}_{j,i}(t)= h^{\mbox{\tiny\text{T}}}_{j,0}(t)\exp\big\{  w_i\gamma^{\mbox{\tiny\text{T}}}_j
+ \mbox{expit}\left\{\eta_{1,i}(t)\right\}\alpha^{\mbox{\tiny\text{T}}}_{j,1}+\eta_{2,i}(t)\alpha^{\mbox{\tiny\text{T}}}_{j,2}+\upsilon_i\alpha^\upsilon_j\big\}$, $j=1,2$. \\
\\
10: & Numerically solve $\exp(-H^{\mbox{\tiny\text{T}}}_{1,i}(t^{\mbox{\tiny\text{T}}\ast}_i))=u^{\mbox{\tiny\text{T}}}_{j,i}$ for $t^{\mbox{\tiny\text{T}}\ast}_{j, i}$~\citep{bender2005generating}, for $j=1,2$, to obtain the individual true event times: $\underset{n\times1}{\textbf{t}_1^{\mbox{\tiny\text{T}}\ast}}$ and $\underset{n\times1}{\textbf{t}_2^{\mbox{\tiny\text{T}}\ast}}$.\\
11: & Calculate the observed event time $t^{\mbox{\tiny\text{T}}}_i=\min(t^{\mbox{\tiny\text{T}}\ast}_{1,i},t^{\mbox{\tiny\text{T}}\ast}_{2,i}, t_{\text{max}})$, where $t_{\text{max}}$ is the deterministic maximum follow-up time: $\underset{n\times1}{\textbf{t}^{\mbox{\tiny\text{T}}}}$.\\
12: & Define the censoring indicator $\delta^{\mbox{\tiny\text{T}}}_i$ as 1 if $t^{\mbox{\tiny\text{T}}}_i = t^{\mbox{\tiny\text{T}}\ast}_{1,i}$, 2 if $t^{\mbox{\tiny\text{T}}}_i = t^{\mbox{\tiny\text{T}}\ast}_{2,i}$, and 0 otherwise. \\
\midrule
\multicolumn{2}{l}{Longitudinal outcome (2/2):} \\
13: & Remove all $\boldsymbol y_{1,i}(t)$ and $\boldsymbol y_{2,i}(t)$ for $t>t_i$. \\
\midrule
\multicolumn{2}{l}{Recurrent outcome:} \\
14: & Generate $n$ random samples from $\mathcal{U}\left(0, 1\right)$, $u^{\mbox{\tiny\text{R}}}_{l,i}$: $\underset{n\times1}{\boldsymbol u^{\mbox{\tiny\text{R}}}_l}$. \\
15: & Define $H^{\mbox{\tiny\text{R}}}_i(t)=\int_{0}^{t} h^{\mbox{\tiny\text{R}}}_i(t) \,ds$, where $h^{\mbox{\tiny\text{R}}}_i(t)= h^{\mbox{\tiny\text{R}}}_0(t)\exp\big\{  w_{i}\gamma^{\mbox{\tiny\text{R}}}
+ \mbox{expit}\left\{\eta_{1,i}(t)\right\}\alpha^{\mbox{\tiny\text{R}}}_1+\eta_{2,i}(t)\alpha^{\mbox{\tiny\text{R}}}_2\big\}$. \\
\\
16: & Numerically solve $\exp(-H^{\mbox{\tiny\text{R}}}_i(t^{\mbox{\tiny\text{R}}\ast}_{l,i}))=u^{\mbox{\tiny\text{R}}}_{l,i}$ for $t^{\mbox{\tiny\text{R}}\ast}_{l, i}$~\citep{bender2005generating}, to obtain the individual true $l$th recurrent event times: $\underset{n\times1}{\textbf{t}_l^{\mbox{\tiny\text{R}}\ast}}$.\\
17: & Calculate the $l$th observed event time $t^{\mbox{\tiny\text{R}}}_{l,i}=\min(t^{\mbox{\tiny\text{R}}\ast}_{l,i},t^{\mbox{\tiny\text{T}}}_i)$: $\underset{n\times1}{\textbf{t}_l^{\mbox{\tiny\text{R}}}}$.\\
18: & Define the censoring indicator as $\delta^{\mbox{\tiny\text{R}}}_{i,l}$ as 1 if $t^{\mbox{\tiny\text{R}}}_{l,i} = t^{\mbox{\tiny\text{T}}}_i$, and 0 otherwise. \\
19: & Repeat steps 14--18 for each individual until $\sum_i t^{\mbox{\tiny\text{R}}}_{l,i} > t^{\mbox{\tiny\text{T}}}_i$. \\
\bottomrule
\end{tabular}
\end{adjustbox}
\end{table}

\begin{table}
\caption{Outline of the data generation process for scenario B.}
\label{tab:sim:jmB}
\centering
\begin{adjustbox}{max width=\textwidth}
\begin{tabular}{lp{16.3cm}}
\toprule
\multicolumn{2}{l}{Longitudinal outcome (1/2):} \\
1: & Generate $n=1000$ random samples from $\mathcal{N}\left(\boldsymbol 0 ,\boldsymbol \Sigma^{-1}\right)$ for the individual-specific random effects, $\boldsymbol b_i$: $\underset{n\times2}{\boldsymbol b}$. \\
2: & Generate $(n\times (n_i-1))$ random samples from $\mathcal{U}\left(0,10\right)$ for the individual visiting times and add the time $0$, $\boldsymbol t_i$: $\underset{(n\cdot n_i)\times1}{\boldsymbol t}$.\\
3: & Generate the $n$ vectors of $n_i$ individual underlying longitudinal responses, $\boldsymbol \mu_i$: $\boldsymbol \mu_i=\mbox{expit}(\boldsymbol \eta_i)$, where $\underset{n_i\times1}{\strut\boldsymbol \eta_i}= \underset{ n_i\times2}{\strut\begin{bmatrix} \boldsymbol 1 & \boldsymbol t_i \end{bmatrix}}\underset{2\times1}{\strut\boldsymbol\beta} + \underset{n_i\times2}{\begin{bmatrix} \boldsymbol 1 & \boldsymbol t_i \end{bmatrix}}\underset{2\times1}{\strut\boldsymbol b_i}$. \\
4: & Generate $(n\times n_i)$ random samples from
$\text{Beta}\left(p,q\right)$, where $p=\phi \times \boldsymbol\mu_i$ and $q=\phi \times (1-\boldsymbol\mu_i)$, for the observed longitudinal responses: $\underset{(n\cdot n_i)\times1}{\boldsymbol y}$.\\
\midrule
\multicolumn{2}{l}{Survival outcome:} \\
7: & Generate $n$ random samples from
$\text{Bern}\left(0.5\right)$ for the individual's group, $w_i$: $\underset{n\times1}{\boldsymbol{w}}$.\\
8: & Generate $n$ random samples from $\mathcal{U}\left(0, 1\right)$, $u_i$: $\underset{n\times1}{\boldsymbol u}$. \\
9: & Define $H_i(t)=\int_{0}^{t} h_i(t) \,ds$, where $h_i(t)= h_0(t)\exp\left\{  w_i\gamma
+ \mbox{expit}\left\{\eta_{1,i}(t)\right\}\alpha^{\mbox{\tiny\text{T}}}_{1,1}\right\}$. \\
\\
10: & Numerically solve $\exp(-H_i(t^\ast_i))=u_i$ for $t^\ast_i$~\citep{bender2005generating} to obtain the individual true event times: $\underset{n\times1}{\textbf{t}^\ast}$.\\
11: & Calculate the observed event times $t_i=\min(t^\ast_i, t_{\text{max}})$, where $t_{\text{max}}$ is the deterministic maximum follow-up time: $\underset{n\times1}{\textbf{t}}$.\\
12: & Define the censoring indicator as $\delta_i=\begin{cases} 1 & t_i \leq t_\text{max},\\ 0 & t_i > t_\text{max}. \end{cases}$\\
\midrule
\multicolumn{2}{l}{Longitudinal outcome (2/2):} \\
13: & Remove all $y_{i}(t)$ for $t>t_i$. \\
\bottomrule
\end{tabular}
\end{adjustbox}
\end{table}

\begin{table}
\caption{Characteristics of the simulated datasets. Abbreviations: ind, individual; IQR, interquartile range; pct, percentile.}
\label{tab:sim:data}
\centering
\begin{adjustbox}{max height=0.47 \textheight}
\begin{tabular}{lrr}
\toprule
& \multicolumn{1}{c}{Scenario A} & \multicolumn{1}{c}{Scenario B} \\ 
\midrule
Number of replicas & 100 & 100 \\
Number of individuals & 1000 & 1000 \\
Number of observations & & \\
\multicolumn{1}{r}{$\text{M}_1$, median (IQR)} & 10945.5 (10852--11087.75) & 15334 (15211.5--15443.25) \\
\multicolumn{1}{r}{$\text{M}_2$, median (IQR)} & 10945.5 (10852--11087.75) & -- \\
& & \\
Number of observations/individual & & \\
\multicolumn{1}{r}{$\text{M}_1$, median, median (IQR)} & 10 (10--10) & 19 (19--19) \\
\multicolumn{1}{r}{$\text{M}_2$, median, median (IQR)} & 10 (10--10) & -- \\
& \raisebox{-\totalheight}{\includegraphics[width=50mm]{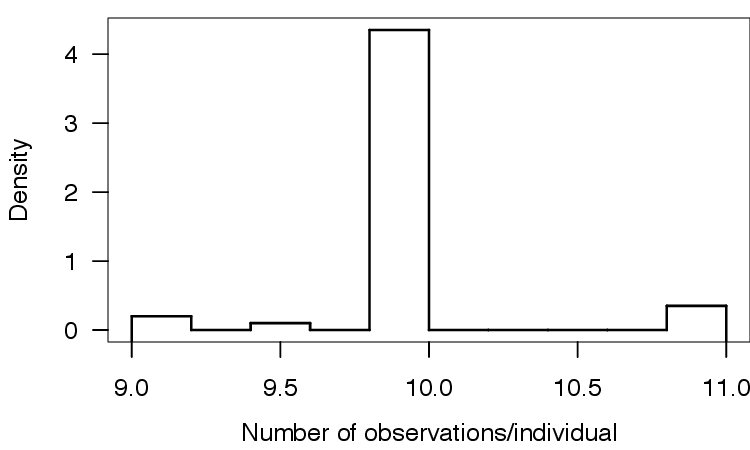}} & \raisebox{-\totalheight}{\includegraphics[width=50mm]{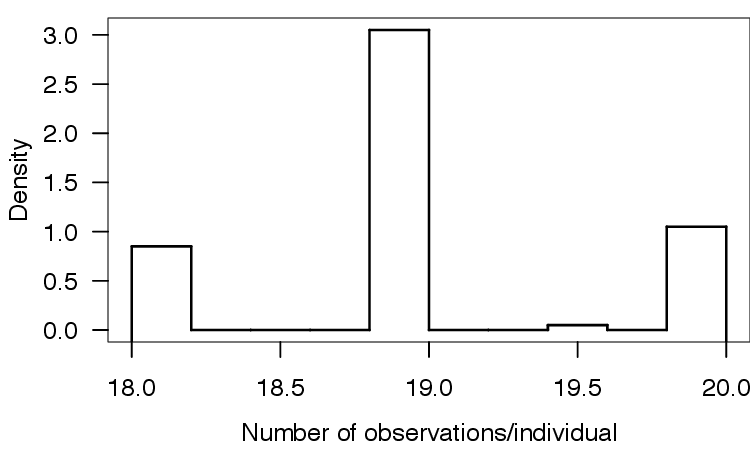}} \\
& & \\
Aggregated follow-up duration & & \\
\multicolumn{1}{r}{$\text{M}_1$, median (IQR)} & 4750.81 (4699.53--4823.44) & 7041.97 (6975.63--7103.06) \\
\multicolumn{1}{r}{$\text{M}_2$, median (IQR)} & 4750.81 (4699.53--4823.44) & -- \\
& & \\
Follow-up duration/individual & & \\
\multicolumn{1}{r}{$\text{M}_1$, median, median (IQR)} & 4.17 (4.09--4.26) & 8.42 (8.31--8.53) \\
\multicolumn{1}{r}{$\text{M}_2$, median, median (IQR)} & 4.17 (4.09--4.26) & -- \\
& \raisebox{-\totalheight}{\includegraphics[width=50mm]{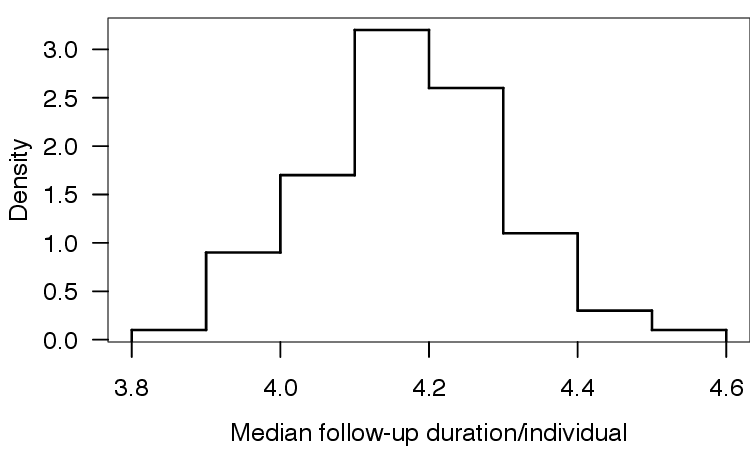}} & \raisebox{-\totalheight}{\includegraphics[width=50mm]{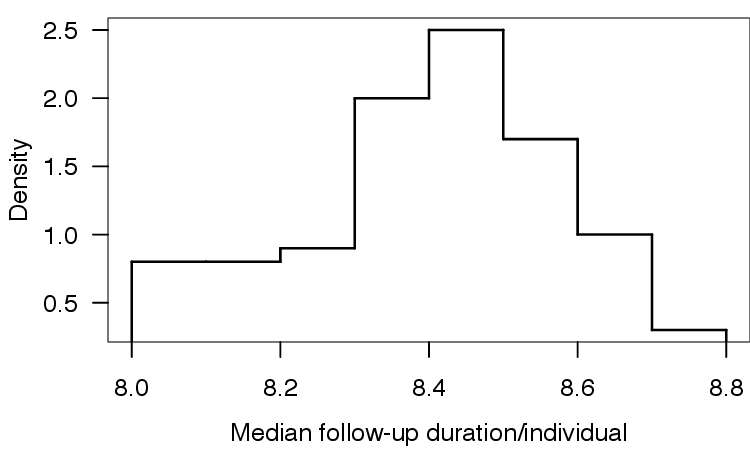}} \\
& & \\
Competing/terminal event time & & \\
\multicolumn{1}{r}{$\text{T}_1$, median, median (IQR)} & 3.97 (3.88--4.07) & 5.48 (5.35--5.58)  \\
\multicolumn{1}{r}{$\text{T}_2$, median, median (IQR)} & 3.97 (3.88--4.07) & -- \\
\multicolumn{1}{r}{Censoring, median, median (IQR)} & 10 (10--10) & 10 (10--10) \\
& \raisebox{-\totalheight}{\includegraphics[width=50mm]{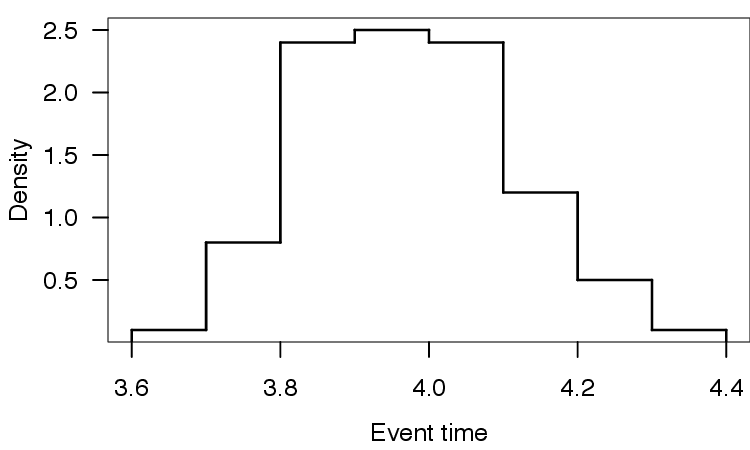}} & \raisebox{-\totalheight}{\includegraphics[width=50mm]{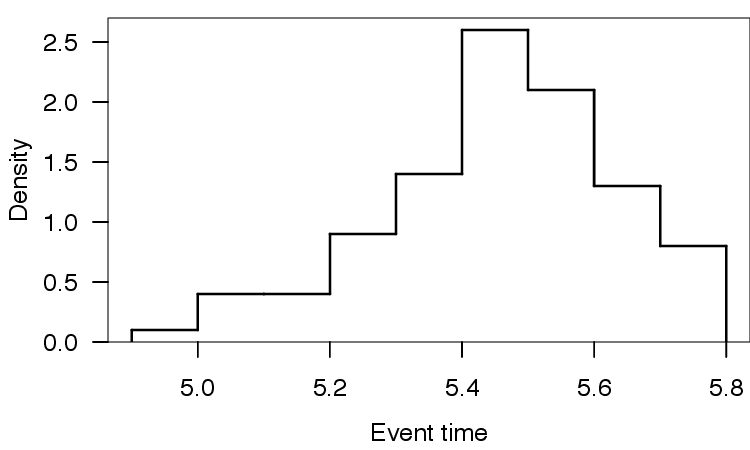}} \\
& & \\
Competing/terminal event & & \\
\multicolumn{1}{r}{$\text{T}_1$, \%, median (IQR)} & 0.42 (0.41--0.43) & 0.54 (0.53--0.55)  \\
\multicolumn{1}{r}{$\text{T}_2$, \%, median (IQR)} & 0.41 (0.4--0.42) & -- \\
\multicolumn{1}{r}{Censoring, \%, median (IQR)} & 0.17 (0.16--0.18) & 0.46 (0.45--0.47)  \\
& & \\
Number of recurrent events/individual & & \\
\multicolumn{1}{r}{Median, median (IQR)} & 3 (3--3) & -- \\
& & \\
Group & & \\
\multicolumn{1}{r}{1, \%, median (IQR)} & 0.5 (0.49--0.51) & 0.5 (0.49--0.51) \\
\bottomrule
\end{tabular}
\end{adjustbox}
\end{table}

\FloatBarrier
\newpage 

\section{CFFPR study}
\label{sec:sup:app}

\begin{figure}
\centerline{\includegraphics[width=0.5\textwidth]{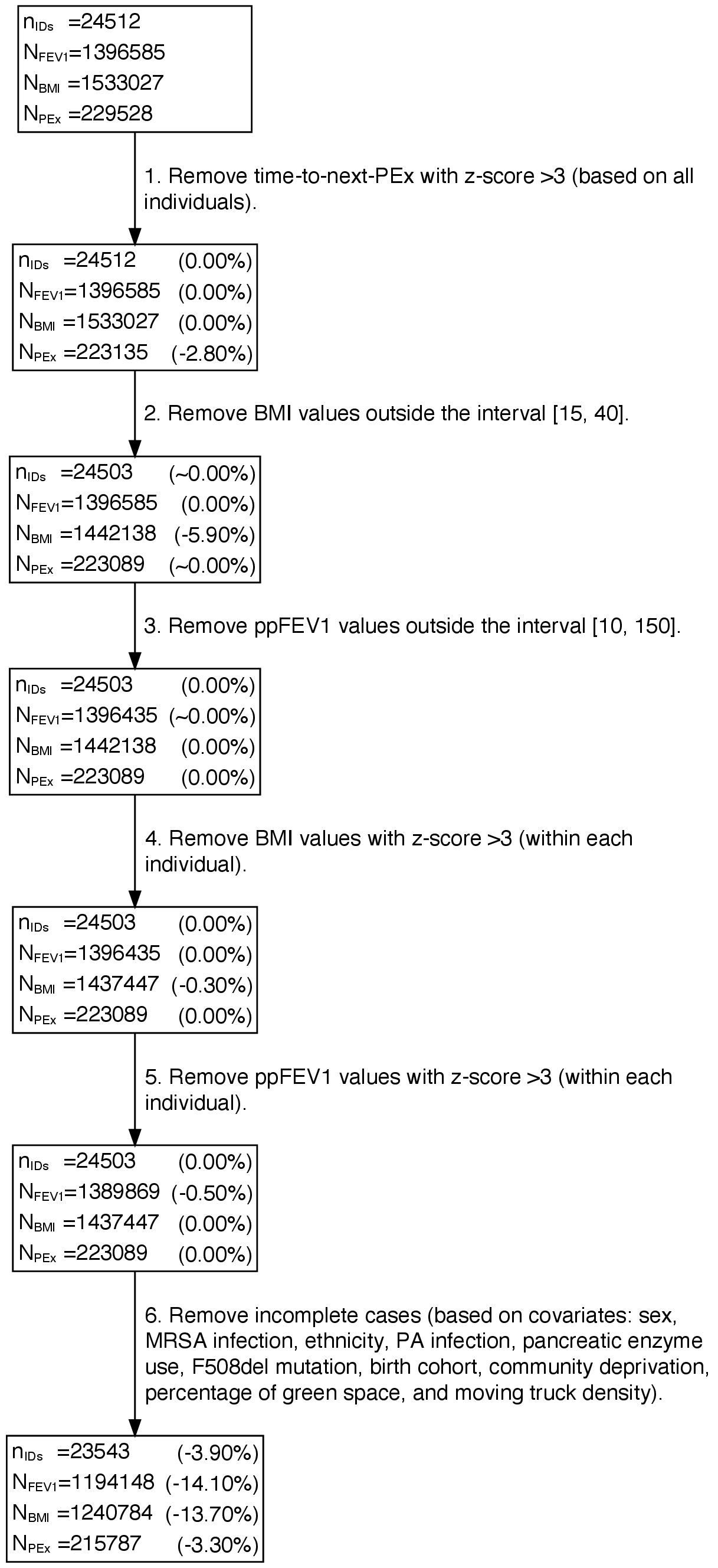}}
\caption{Cleaning process of CFFPR data sample. The sequence of steps 1--5 was employed to identify and remove atypical measurements likely arising from data entry errors. n\textsubscript{IDs}: number of individuals; n\textsubscript{FEV1}: number of ppFEV\textsubscript{1} measurements; n\textsubscript{BMI}: number of BMI measurements; n\textsubscript{PEx}: number of PEx.}
\label{fig:app:incexc}
\end{figure}

\begin{table}
\caption{Follow-up, demographic, social, and clinical characteristics of the CF individuals analyzed. Abbreviations: BMI, body mass index; CF, cystic fibrosis; IQR, interquartile range; PEx, pulmonary exacerbation; ppFEV\textsubscript{1}, percent predicted forced expiratory volume in one second. \textsuperscript{\textdagger}: Percentage of greenspace, impervious, and tree canopy areas within the ZIP Code Tabulation Area (ZCTA) derived from the National Land Cover Database~\citep{jin2019overall}}
\label{tab:app:data}
\centering
\begin{adjustbox}{max height=0.43\textheight}
\begin{tabular}{lr}
\toprule
Characteristics & \\
\midrule
Number of individuals & 23,543 \\
Number of measurements &  \\
\multicolumn{1}{r}{ppFEV\textsubscript{1}} & 1,523,40 \\
\multicolumn{1}{r}{BMI} & 1,523,40 \\
Number of measurements/individual & \\
\multicolumn{1}{r}{ppFEV\textsubscript{1}, median (IQR)} & 46.00 (27.00--69.00) \\
\multicolumn{1}{r}{BMI, median (IQR)} & 48.00 (27.00--72.00) \\
Aggregated follow-up duration (years) &  \\
\multicolumn{1}{r}{ppFEV\textsubscript{1}} & 266,345.20 \\
\multicolumn{1}{r}{BMI} & 262,875.00 \\
Follow-up duration/individual (years) & \\
\multicolumn{1}{r}{ppFEV\textsubscript{1}, median (IQR)} & 11.92 (6.97--16.76) \\
\multicolumn{1}{r}{BMI, median (IQR)} & 11.72 (6.85--16.61) \\
Baseline age (years) & \\
\multicolumn{1}{r}{ppFEV\textsubscript{1}, median (IQR)} & 11.37 (6.36--20.19) \\
\multicolumn{1}{r}{BMI, median (IQR)} & 11.80 (6.66--20.15) \\
Age at end of follow-up (years) & \\
\multicolumn{1}{r}{Censoring, median (IQR)} & 23.50 (17.07--32.15) \\
\multicolumn{1}{r}{Lung transplantation, median (IQR)} & 28.52 (22.84--36.55) \\
\multicolumn{1}{r}{Death, median (IQR)} & 26.57 (21.36--35.93) \\
Competing terminal event & \\
\multicolumn{1}{r}{Censoring} & 16,751 (71.15\%) \\
\multicolumn{1}{r}{Lung transplantation} & 2,562 (10.88\%) \\
\multicolumn{1}{r}{Death} & 4,230 (17.97\%) \\
Number of PEx/individual & \\
\multicolumn{1}{r}{Median (IQR)} & 7.00 (3.00--14.00) \\
Interval between consecutive PEx (years) & \\
\multicolumn{1}{r}{Median (IQR)} & 0.34 (0.15--0.77) \\
Baseline ppFEV\textsubscript{1} & \\
\multicolumn{1}{r}{Median (IQR)} & 80.30 (59.70--95.90) \\
Baseline BMI & \\
\multicolumn{1}{r}{Median (IQR)} & 17.17 (15.66--20.31) \\
Birth cohort & \\
\multicolumn{1}{r}{$<$1993} & 13,895 (59.02\%) \\
\multicolumn{1}{r}{[1993, 1998)} & 3,672 (15.60\%) \\
\multicolumn{1}{r}{$\geq$1998} & 5,976 (25.38\%) \\
Genotype (F508del) & \\
\multicolumn{1}{r}{Homozygous} & 11,236 (47.73\%) \\
\multicolumn{1}{r}{Heterozygous} & 8,655 (36.76\%) \\
\multicolumn{1}{r}{Neither} & 3,652 (15.51\%) \\
Sex & \\
\multicolumn{1}{r}{Female} & 11,829 (50.24\%) \\
Ethnicity & \\
\multicolumn{1}{r}{Hispanic} & 1,767 (7.51\%) \\
\multicolumn{1}{r}{Other} & 21,776 (92.49\%) \\
Neighborhood deprivation index & \\
\multicolumn{1}{r}{Median (IQR)} & 0.33 (0.27--0.40) \\
Percentage of green space\textsuperscript{\textdagger} & \\
\multicolumn{1}{r}{Median (IQR)} & 89.81 (71.81--96.94) \\
Moving-truck density (truck-meters/m\textsuperscript{2}) & \\
\multicolumn{1}{r}{Median (IQR)} & 0.18 (0.00--0.94) \\
Pancreatic enzymes intake & \\
\multicolumn{1}{r}{At baseline} & 6,887 (29.25\%) \\
\multicolumn{1}{r}{Throughout follow-up} & 1,868 (7.93\%) \\
\multicolumn{1}{r}{Sometime during follow-up} & 22,564 (95.84\%) \\
\bottomrule
\end{tabular}
\end{adjustbox}
\end{table}

Here, we explain how to interpret the association parameter estimates from the proposed joint model. 

A $q$-unit increase in the ppFEV\textsubscript{1}'s expected value changes the hazard rate of PEx by a factor of $\exp\left\{\frac{\alpha^{\mbox{\tiny\text{R}}}_{1,1}}{150}\times q\right\}$. We divide $\alpha^{\mbox{\tiny\text{R}}}_{1,1}$ by $150$ to rescale the parameter to the original marker scale. The same rationale applies to the association parameters $\alpha^{\mbox{\tiny\text{T}}}_{1,1,1}$ and $\alpha^{\mbox{\tiny\text{T}}}_{2,1,1}$ regarding the risks of transplantation and death, respectively. 

\sloppy A $p$-unit increase in the rate of decline of ppFEV\textsubscript{1}'s expected value changes the hazard rate of transplantation by a factor of $\exp\left\{\frac{\alpha^{\mbox{\tiny\text{T}}}_{1,1,2}}{150}\times p\right\}$. We divide $\alpha^{\mbox{\tiny\text{T}}}_{1,1,2}$ by $150$ because we transform the marker from the range of 0 to 150 to the desired range of 0 to 1, in which the beta distribution is defined. In other words, $\text{ppFEV}_{1\,i}(t)=\frac{\text{ppFEV}^\ast_{1\,i}(t)}{150}$, where $\text{ppFEV}^\ast_{1\,i}(t)$ and $\text{ppFEV}_{1\,i}(t)$ denote markers in the original and transformed scales, respectively. Given that $\frac{\text{d}}{\text{d}t}\,\text{ppFEV}_{1\,i}(t)=\frac{1}{150}\,\frac{\text{d}}{\text{d} t}\,\text{ppFEV}^\ast_{1\,i}(t)$ and noting that $\alpha^{\mbox{\tiny\text{T}}}_{1,1,2}\left(\frac{1}{150}\,\frac{\text{d}}{\text{d} t}\,\text{ppFEV}^\ast_{1\,i}(t)\right)=\frac{\alpha^{\mbox{\tiny\text{T}}}_{1,1,2}}{150}\left(\frac{\text{d}}{\text{d} t}\,\text{ppFEV}^\ast_{1\,i}(t)\right)$, we can obtain the association paramter in the original scale by doing $\alpha^{\mbox{\tiny\text{T}}*}_{1,1,2}=\frac{\alpha^{\mbox{\tiny\text{T}}}_{1,1,2}}{150}$. The same reasoning applies to the association parameter $\alpha^{\mbox{\tiny\text{T}}}_{2,1,2}$ regarding the risks of death.

An $b$-unit increase in the BMI's standardized cumulative value changes the hazard rate of PEx by a factor of $\exp\left\{\alpha^{\mbox{\tiny\text{R}}}_{2,1}\times b\right\}$. The same reasoning applies to the association parameters $\alpha^{\mbox{\tiny\text{T}}}_{1,2,1}$ and $\alpha^{\mbox{\tiny\text{T}}}_{2,2,1}$ regarding the risks of transplantation and death, respectively. 

An $f$-unit increase in the frailty term $\upsilon^{\mbox{\tiny\text{R}}}_i$ changes the hazard rate of transplantation by a factor of $\exp\left\{\alpha^\upsilon_{1}\times f\right\}$. The same reasoning applies to the association parameter $\alpha^\upsilon_{2}$ for the risk of death.

\end{document}